\documentclass{jfm}

\usepackage{graphicx}
\usepackage{newtxtext}
\usepackage{newtxmath}
\usepackage{natbib}
\usepackage{hyperref}
\hypersetup{
    colorlinks = true,
    urlcolor   = blue,
    citecolor  = black,
}

\newcommand{\RomanNumeralCaps}[1]
\linenumbers

\usepackage{subcaption}
\usepackage{diagbox}
\usepackage{makecell}

\newcommand{\rme}{\mathrm{e}}
\newcommand{\rmi}{\mathrm{i}}
\newcommand{\rmexp}{\mathrm{exp}}

\usepackage{amsmath}
\DeclareMathOperator{\sign}{sign}

\usepackage{soul}
\usepackage[normalem]{ulem}
\newcommand\redsout{\bgroup\markoverwith{\textcolor{red}{\rule[0.2ex]{2pt}{2pt}}}\ULon}

\usepackage{xcolor}

\title{Internal shear layers generated by a vertically oscillating cylinder in unbounded and bounded rotating fluids} 

\author{Jiyang He\aff{1,2}
  \corresp{\email{jiyanghe123@gmail.com}},
  Benjamin Favier\aff{2}
 \and St\'ephane Le Diz\`es\aff{2}}

\affiliation{
\aff{1}Department of Ocean Science, The Hong Kong University of Science and Technology, Hong Kong, China
\aff{2}Aix Marseille Univ, CNRS, Centrale Med, IRPHE, Marseille, France
}

\begin{document}
\maketitle

\begin{abstract}
In rotating fluids, the viscous smoothing of inviscid singular inertial waves leads to the formation of internal shear layers.
 In previous works, we analysed the internal shear layers excited by a viscous forcing (longitudinal libration) in a spherical shell geometry (He \textit{et al.}, \textit{J. Fluid Mech.} {\bf 939}, A3, 2022; {\bf 974}, A3, 2023). We now consider the stronger inviscid forcing corresponding to the vertical oscillation of the inner boundary. 
 We limit our analysis to two-dimensional geometries but examine three different configurations: 
freely-propagating wave beams in an unbounded domain and two wave patterns (a periodic orbit and an attractor)  in a cylindrical shell geometry. 
The asymptotic structures of the internal shear layers are assumed to follow the similarity solution of Moore \& Saffman (\textit{Phil. Trans. R. Soc. Lond. A}, 264 (1156), 1969, 597-634) in the small viscous limit.
The two undefined parameters of the similarity solution (singularity strength and amplitude) are derived by asymptotically matching the similarity solution with the inviscid solution.
For each case, the derivation of the latter is achieved either through separation of variables combined with analytical continuation or the method of characteristics.
Global inviscid solutions, when obtained, closely match numerical solutions for small Ekman numbers far from the critical lines, while viscous asymptotic solutions show excellent performance near those lines.
The amplitude scalings of the internal shear layers excited by an inviscid forcing are found to be divergent as the Ekman number $E$ decreases, specifically $O(E^{-1/6})$ for the critical point singularity and $O(E^{-1/3})$ for attractors, 
in contrast to the convergent scalings found for a viscous forcing.

\end{abstract}


\section{Introduction}
Inertial waves are ubiquitous in rotating fluids. They are associated with the Coriolis  restoring force.  
They propagate at a fixed angle $\pi/2-\theta_c$ relative to the rotation axis that only depends on the inertial wave frequency $\omega^*$ through the dispersion relation $\omega^*= 2 \Omega^* \cos\theta_c$ ($\Omega^*$ being the rotation rate) \citep{greenspan1968theory}. This angle is conserved when waves reflect on boundaries.
Owing to this property, the width of an inertial wave beam contracts or expands, depending on the inclination of the boundary relative to the propagation angle $\theta_c$ and the rotation axis\citep{phillipsEnergyTransferRotating1963,ledizesReflectionOscillatingInternal2020}.
The contraction effect is responsible for the creation of inviscid singularities on the critical characteristics, such as the tangent lines from the critical points on a convex boundary \citep{ledizesCriticalSlopeSingularities2024}, and the attractors inside a bounded domain \citep{ogilvie2020internal}.
Inviscid singularities can be understood as the outcome of the hyperbolic character of the Poincar\'e equation, namely, the ability to transport singularities \citep{kerswellInternalShearLayers1995} and to form an ill-posed Cauchy problem in closed domains \citep{Rieutord2000}.

In numerical simulations and experiments, these inviscid singularities are smoothed out by viscosity, giving rise to oscillatory internal shear layers around the critical characteristics \citep{kerswellInternalShearLayers1995}.
These layers become the dominant feature of a global viscous solution when inviscid singularities are present.
The internal shear layers associated with a critical slope singularity have been experimentally observed in both rotating \citep{greenspan1968theory} and 
stratified \citep{Mowbray1967,zhangExperimentalStudyInternal2007} fluids, owing to the similarity between the dispersion relation of inertial waves and that of internal gravity waves.  
Similarly in a bounded domain, attractors give rise to internal shear layers that have been observed in various configurations \citep{maasObservationInternalWave1997,Hazewinkel2008,Klein2014}.

The mathematical description of the internal oscillatory shear layers relies on local asymptotic analysis in the limit of the small Ekman number $E$ (defined as the ratio between the viscous term and the Coriolis term, see below).
Although different scalings have been identified for the width of these layers in forced and eigenvalue problems, including $E^{1/3}$, $E^{1/4}$, $E^{1/5}$ and $E^{1/6}$ \citep{kerswellInternalShearLayers1995,rieutordAxisymmetricInertialModes2018}, the scaling in $E^{1/3}$ appears as the natural scaling in a forced problem for 
the width of the
 internal shear layer associated with a critical slope singularity or an attractor. 
For the former, the scaling was derived by dominant balance \citep{waltonWavesThinRotating1975,waltonViscousShearLayers1975,kerswellInternalShearLayers1995}, and numerically validated \citep{linLibrationdrivenInertialWaves2020}.
For the latter, it was inferred by \citet{ogilvieWaveAttractorsAsymptotic2005} and numerically validated by \citet{grisouardNumericalSimulationTwodimensional2008}.

The viscous structure of these layers has been the subject of many works. 
The first theoretical work in an oscillatory context goes probably back to \citet{waltonWavesThinRotating1975}. 
The building block is now believed to be a class of similarity solutions that was introduced by \citet{mooreStructureFreeVertical1969} in rotating fluids and  \citet{thomasSimilaritySolutionViscous1972} in stratified fluids.
These solutions depend on an index $m$ that is related to the strength of the underlying inviscid singularity \citep{ledizesCriticalSlopeSingularities2024}.
The similarity solutions were initially used to describe in the far field the solution  generated by  a localized oscillating source \citep{hurleyGenerationInternalWaves1997,voisinLimitStatesInternal2003}. In this context, \citet{machicoaneInfluenceMultipoleOrder2015} have shown that the index $m$ is linked to multipolar character of the source. \citet{Tilgner2000} and \cite{LeDizes2015} have also shown that they  describe the thin shear layers generated from the border of an oscillating disk. 
The use of the similarity solutions to describe the oscillatory internal shear layers generated from a critical point  was first made by \citet{ledizesInternalShearLayers2017} in the case of a viscous forcing for a librating object 
in an unbounded domain.
\citet{heInternalShearLayers2022,heInternalShearLayers2023} extended these results to a bounded geometry  for both a periodic wave pattern and an attractor. 
For the attractor case, they  used original ideas that were introduced in an inclined rotating square subject to a body forcing by  \citet{ogilvieWaveAttractorsAsymptotic2005}.  
These ideas will  also be  used in the present study.

In our previous works \citep{ledizesInternalShearLayers2017,heInternalShearLayers2022,heInternalShearLayers2023}, the internal shear layers were generated by libration, that is  by a viscous forcing at the critical point. 
In the present study, we consider an inviscid forcing, specifically vertical oscillation, where internal shear layers are forced through pressure coupling.
For such a forcing, the nature of the internal shear layers generated from a critical point is expected to be different.  \cite{ledizesCriticalSlopeSingularities2024} has explained that in contrast to the viscous case, the amplitude of the internal shear layer cannot be obtained in closed form from a local analysis of the critical point.  
Our analysis will therefore go through the derivation of global inviscid solutions before considering  internal shear layers. 
It will be limited to two-dimensional geometries.


We start with the classical problem of an oscillating cylinder in an unbounded domain. 
The inviscid problem has been solved  in a stratified fluid by \citet{hurleyGenerationInternalWaves1997a}. 
We provide here the inviscid solution in a rotating fluid using the same approach. 
This solution is singular along the characteristics issued from the critical points. 
From the behavior of the solution close to the singularity, we obtain the index and the amplitude of the similarity solution that describes the viscous structure 
of the internal shear layer.  
 An alternative method 
based on a global viscous solution is also given in Appendix~\ref{app:global_viscous_solution}. 
The approximation that is obtained is compared with viscous numerical results obtained for very small Ekman numbers. 
We then consider the cylindrical shell geometry. 
The ray patterns strongly depend on the frequency. 
For a frequency $\omega^*= \sqrt{2}\,\Omega^*$, the propagation is at 45 degrees with respect to the rotation axis.
We  first consider this case for which the ray trajectories are periodic. 
A global inviscid solution is obtained by writing down  the 
conditions of reflection on each trajectory.  
This solution is singular on the critical lines.
To obtain an approximation valid close to these lines, we  adapt the method used in \citet{heInternalShearLayers2022} for a viscous forcing. 
Both the index and the amplitude of the similarity solutions needed to build this asymptotic solution are obtained by matching this solution with the global inviscid solution.  Again, the theoretical 
results are compared with numerical results. 
In the last section, we consider the case of an attractor.  
An asymptotic solution is obtained  close to the attractor using the analysis developed in   \citet{heInternalShearLayers2023}, and then compared to numerical results. 
 Additional numerical results for particular attractors for which there is no asymptotic theory are also provided
in Appendix~\ref{app:attractors}.
The paper is concluded by summarizing the main results, emphasizing the difference between inviscid and viscous forcings, and discussing the applicability of the 
results in a three-dimensional context. 
 


\section{Framework}

\subsection{Geometry and governing equations}
\label{sec:configurations}
\begin{figure}
    \centering
    \begin{subfigure}{0.45\textwidth}
    \centering
        \includegraphics[width=\textwidth]{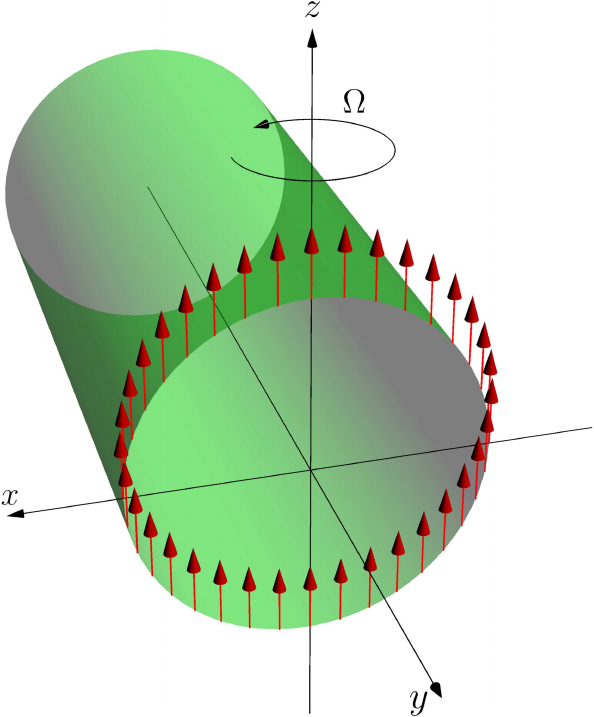}
        \caption{A cylinder oscillating vertically in an unbounded rotating fluid.}
        \label{fig:config_cylinder_vertical_oscillation_unbounded_domain}
    \end{subfigure} 
    ~~~
    \begin{subfigure}{0.45\textwidth}
    \centering
        \includegraphics[width=\textwidth]{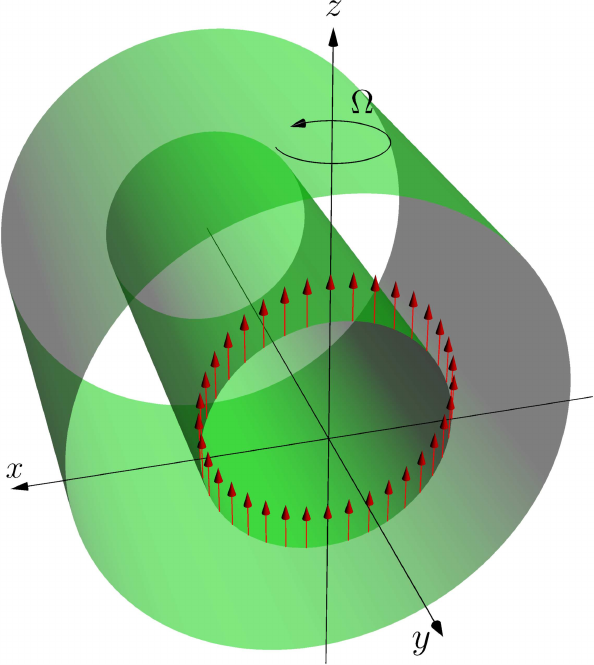}
        ~\vspace{0.3cm}
        \caption{A cylindrical shell domain of rotating fluid with an inner cylinder oscillating vertically.}
        \label{fig:config_cylinder_vertical_oscillation}
    \end{subfigure}
    \caption{Two dimensional configurations in unbounded $(a)$ and bounded $(b)$ domains.}
    \label{fig:config}
\end{figure}

We consider the internal shear layers generated by a two-dimensional cylindrical body oscillating vertically in either an unbounded or bounded domain.
For the former, the configuration is a circular cylinder immersed in an unbounded fluid, as shown in figure \ref{fig:config_cylinder_vertical_oscillation_unbounded_domain}; for the latter, the geometry is a cylindrical annulus, as shown in figure \ref{fig:config_cylinder_vertical_oscillation}.
In both cases, the cylinders' axes   are horizontal and extend to infinity, making the problems two-dimensional.
We adopt a Cartesian coordinate system, where the axis $Oz$ is vertical to the horizontal plane $Oxy$.
The  cylinders' axes  are along the axis $Oy$ such that cross-sections lie in the vertical plane $Oxz$.
The flow is that of an incompressible fluid with constant kinematic viscosity $\nu^*$, rotating around the axis $Oz$ with a uniform rotation rate $\Omega^*$.
Although the flow field  only depends  on $x$ and $z$, it has three velocity components.

Due to the two dimensional nature of the problem, we can use a polar coordinate system to describe the flow field, with the radial coordinate $\varrho=\sqrt{x^2+z^2}$ and angular coordinate $\vartheta=\arctan{(z/x)}$.
 The flow domain is defined by $\varrho \in[\varrho_i^*, \varrho_o^*]$, where $\varrho_o^*$ is infinite for the unbounded case.
In the bounded case, we write $[\varrho_i^*, \varrho_o^*]=[\eta\varrho^*,\varrho^*]$,  with $\eta \in]0,1[$ defining the aspect ratio. 
For convenience of comparison, the radial range of the unbounded domain is chosen as $[\varrho_i^*, \varrho_o^*]=[\eta\varrho^*,+\infty]$, with the same inner radius as the bounded domain.

For both  configurations, lengths are non-dimensionalised by $\varrho^*$, resulting in non-dimensional radial ranges of $[\eta,1]$ for the bounded domain and $[\eta,+\infty]$ for the unbounded domain.
Thus, the non-dimensional radius of the forced cylinder is $\eta$ in both cases. This choice of non-dimensionalisation is made to facilitate the comparison between  inviscid results in unbounded and bounded domains.
Time is non-dimensionalised by the angular period $1/\Omega^*$.
Using these scales, the Ekman number is defined as
\begin{equation}
    E=\frac{\nu^*}{\Omega^*\varrho^{*2}}.
\end{equation}

The imposed harmonic forcing is the vertical oscillation of the cylinder in the unbounded domain or the inner cylinder in the bounded domain (see red arrows in figure \ref{fig:config}), with the displacement amplitude $\varepsilon=\varepsilon^*/\varrho^*$ ($\varepsilon\ll 1$) and the frequency $\omega=\omega^*/\Omega^*$.
As in our previous works \citep{ledizesInternalShearLayers2017,heInternalShearLayers2022,heInternalShearLayers2023}, we focus on the linear harmonic response 
and consider solutions in the form
\begin{equation}
    (\boldsymbol{v},p)\rme^{-\rmi\omega t}+c.c.,
\end{equation}
where $c.c.$ denotes the complex conjugate.
The velocity $\boldsymbol{v}$ and pressure $p$ satisfy the linearised incompressible Navier-Stokes equations in the rotating frame:
\begin{subeqnarray}\label{eq:governing_equations}
    -\rmi\omega v_x-2v_y+\frac{\partial p}{\partial x}-E\nabla^2v_x=0, \\[3pt]
    -\rmi\omega v_z+\frac{\partial p}{\partial z}-E\nabla^2v_z=0, \\[3pt]
    -\rmi\omega v_y+2v_x-E\nabla^2v_y=0, \\[3pt]
    \frac{\partial v_x}{\partial x}+\frac{\partial v_z}{\partial z}=0,
\end{subeqnarray}
with the Laplacian operator
\begin{equation}
    \nabla^2=\partial^2/\partial x^2+\partial^2/\partial z^2.
\end{equation}

As the oscillations are assumed to be very small ($\varepsilon \ll 1$), the (inner) cylinder can be assumed fixed at leading order. The boundary conditions  at the surface of the (inner) cylinder read 
\begin{equation}\label{eq:boundary_condition}
    \boldsymbol{v}=\boldsymbol{e}_z, \quad \mbox{at} \quad \varrho=\sqrt{x^2+z^2}=\eta.
\end{equation}
We apply no-slip boundary conditions at the outer cylinder. In the unbounded case, we apply a condition of radiation which states that the field should be composed of outgoing waves at infinity. 
In a viscous fluid, this condition is equivalent to the vanishing of the solution at infinity. In the numerical code, the condition of radiation will be applied by damping 
the waves propagating after a large but finite radial distance. 

We shall confine ourselves to the inertial wave regime, namely, $0<\omega<2$, within which the dominant structures of the linear response are internal shear layers.

As shown in figure \ref{fig:config}, the imposed vertical oscillation (\ref{eq:boundary_condition}) is anti-symmetric and symmetric about the horizontal axis $Ox$ and vertical axis $Oz$ respectively.
These symmetries, in turn, constrain the symmetries of the directly forced velocity components $v_x$ and $v_z$.
Mathematically, $v_x$ satisfies
\refstepcounter{equation}\label{eq:symmetry_vx}
$$
v_x(\varrho,-\vartheta)=-v_x(\varrho,\vartheta); \quad v_x(\varrho,\pi-\vartheta)=-v_x(\varrho,\vartheta), \eqno{(\theequation{\mathit{a},\mathit{b}})}
$$
and $v_z$ satisfies 
\refstepcounter{equation}\label{eq:symmetry_vz}
$$
v_z(\varrho,-\vartheta)=v_z(\varrho,\vartheta); \quad v_z(\varrho,\pi-\vartheta)=v_z(\varrho,\vartheta). \eqno{(\theequation{\mathit{a},\mathit{b}})}
$$
The symmetries of the transverse field $v_y$ are not as straightforward.
We note however that $v_y$ should satisfy the same mathematical relations as $v_x$, due to the governing equation (\ref{eq:governing_equations}$c$).
Thus, $v_y$ satisfies the following relations that are analogous to (\ref{eq:symmetry_vx})
\refstepcounter{equation}\label{eq:symmetry_vy}
$$
v_y(\varrho,-\vartheta)=-v_y(\varrho,\vartheta); \quad v_y(\varrho,\pi-\vartheta)=-v_y(\varrho,\vartheta). \eqno{(\theequation{\mathit{a},\mathit{b}})}
$$
Since $v_y$ is perpendicular to the $Oxz$ plane, it is anti-symmetric about the two axes $Ox$ and $Oz$.

\subsection{Numerical method}
\label{sec:numerical_method}
The numerical method from our previous work \citep{heInternalShearLayers2023} is adopted here.
The governing equations (\ref{eq:governing_equations}) are numerically solved in polar coordinates $(\varrho, \vartheta)$.
In terms of the streamfunction $\psi$ and the associated variable $\chi$,
\refstepcounter{equation}
$$
v_\varrho=-\frac{1}{\varrho}\frac{\p\psi}{\p\vartheta}, \quad v_\vartheta=\frac{\p\psi}{\p\varrho}, \quad v_y=\chi, \eqno{(\theequation{\mathit{a}-\mathit{c}})}
$$
the governing equations (\ref{eq:governing_equations}) are recast to
\begin{subeqnarray}\label{eq:governing_equation_for_streamfunction}
    -\rmi\omega\nabla^2\psi+2\left(\sin\vartheta\frac{\p\chi}{\p\varrho}+\frac{\cos\vartheta}{\varrho}\frac{\p\chi}{\p\vartheta}\right)-E\nabla^4\psi=0,\\[3pt]
    -\rmi\omega\chi-2\left(\sin\vartheta\frac{\partial\psi}{\partial\varrho}+\frac{\cos\vartheta}{\varrho}\frac{\p\psi}{\p\vartheta}\right)-E\nabla^2\chi=0,
\end{subeqnarray}
with the operator 
\begin{equation}
    \nabla^2=\frac{\p^2}{\p\varrho^2}+\frac{1}{\varrho}\frac{\p}{\p\varrho}+\frac{1}{\varrho^2}\frac{\p^2}{\p\vartheta^2}.
\end{equation}

The streamfunction $\psi$ and the associated variable $\chi$ are expanded by Fourier series in the angular direction as
\refstepcounter{equation}\label{eq:fourier_expansion}
$$
\psi=\sum_{l=-\infty}^{+\infty}\psi_l(\varrho)\rme^{\rmi l\vartheta}, \quad 
\chi=-\rmi\sum_{l=-\infty}^{+\infty}\chi_l(\varrho)\rme^{\rmi l\vartheta}.
\eqno{(\theequation{\mathit{a},\mathit{b}})}
$$
The projection of the governing equations (\ref{eq:governing_equation_for_streamfunction}) onto this basis leads to
\begin{subeqnarray}\label{eq:spectral_equations}
    \rmi\omega\nabla_l^2\psi_l+(\chi_{l-1}^\prime-\chi_{l+1}^\prime)-\frac{1}{\varrho}\left[(l-1)\chi_{l-1}+(l+1)\chi_{l+1}\right]+E\nabla_l^4\psi_l=f^{\psi}_l, \\[3pt]
    \rmi\omega\chi_l+(\psi_{l-1}^\prime-\psi_{l+1}^\prime)-\frac{1}{\varrho}\left[(l-1)\psi_{l-1}+(l+1)\psi_{l+1}\right]+E\nabla_l^2\chi_l=f^{\chi}_l,
\end{subeqnarray}
with
\begin{equation}
    \nabla_l^2=\frac{\mathrm{d}^2}{\mathrm{d}\varrho^2}+\frac{1}{\varrho}\frac{\mathrm{d}}{\mathrm{d}\varrho}-\frac{l^2}{\varrho^2}.
\end{equation}
The right-hand side terms $f^{\psi}_l$ and $f^{\chi}_l$ represent a sponge layer absorbing outgoing waves and will be specified later.

Not all Fourier components in the expansion (\ref{eq:fourier_expansion}) are necessary in our particular case.
From the symmetry properties of the velocity fields (\ref{eq:symmetry_vx}-\ref{eq:symmetry_vy}), $\psi$ and $\chi$ should satisfy
\refstepcounter{equation}
$$
\psi(-\vartheta)=\psi(\vartheta); \quad \psi(\pi-\vartheta)=-\psi(\vartheta) \eqno{(\theequation{\mathit{a},\mathit{b}})}
$$
and
\refstepcounter{equation}
$$
\chi(-\vartheta)=-\chi(\vartheta); \quad \chi(\pi-\vartheta)=-\chi(\vartheta) \eqno{(\theequation{\mathit{a},\mathit{b}})}
$$
respectively.
The Fourier expansion (\ref{eq:fourier_expansion}) can thus be reduced to
\refstepcounter{equation}
$$
\psi=2\sum_{l'=1}^{+\infty}\psi_{2l'-1}(\varrho)\cos{[(2l'-1)\vartheta]}, \quad \chi=2\sum_{l'=1}^{+\infty}\chi_{2l'}(\varrho)\sin{(2l'\vartheta)}, \eqno{(\theequation{\mathit{a},\mathit{b}})}
$$
where only Fourier components $\psi_1,\chi_2,\psi_3,\chi_4,\cdots$ with positive $l$ are solved.

The spectral equations (\ref{eq:spectral_equations}) are supplemented with boundary conditions at the inner and outer boundaries of the radial domain.
The vertical oscillation (\ref{eq:boundary_condition}) at the inner boundary yields the following boundary condition
\begin{equation}
    \psi_l=\frac{\varrho}{2}\delta_{1l}, \quad \frac{\mathrm{d}\psi_l}{\mathrm{d}\varrho}=\frac{1}{2}\delta_{1l}, \quad \chi_l=0 \quad \mbox{at} \quad \varrho=\eta.
\end{equation}
The boundary condition at the outer boundary is
\begin{equation}
    \psi_l=\frac{\mathrm{d}\psi_l}{\mathrm{d}\varrho}=\chi_l=0 \quad \mbox{at} \quad \varrho= 1 \quad \mbox{or} \quad \varrho\rightarrow +\infty,
\end{equation}
for the bounded or unbounded domain respectively.

Special treatment is required for the unbounded domain, which is truncated at a finite radius.
The position of the outer boundary is chosen to be far from the source, within the limit of numerical resources.
A sponge layer is added near the outer boundary to absorb all outgoing waves.
We directly implement the sponge layer in the spectral equations (\ref{eq:spectral_equations}) through the right-hand terms $f^\psi_l$ and $f^\chi_l$.
They take the form of a damping function as follows
\begin{subeqnarray}\label{eq:sponge_layer}
    f^\psi_l=\frac{1}{2\Upsilon}\left[1+\tanh{\left(\frac{\varrho-\varrho_s}{\Lambda}\right)}\right]\nabla_l^2\psi_l,\\[3pt]
    f^\chi_l=\frac{1}{2\Upsilon}\left[1+\tanh{\left(\frac{\varrho-\varrho_s}{\Lambda}\right)}\right]\chi_l.
\end{subeqnarray}
The parameter $\Upsilon$ is the time scale at which waves are damped and is simply taken as $\omega^{-1}$.
The parameter $\varrho_s$ is the position where the sponge layer is centred and should be close to the outer boundary.
The parameter $\Lambda$ controls the domain range affected by the sponge layer.
Suitable values of $\varrho_s$ and $\Lambda$ are chosen by trial and error.

Finally, the spectral equations (\ref{eq:spectral_equations}) are truncated at the Fourier order $L$ and further discretized using the Chebyshev collocation method, with the derivatives relative to $\varrho$ replaced by the Chebyshev differentiation matrices of order $N+1$.
The resulting block tridiagonal system is solved by the block tridiagonal algorithm as described in \citet{heInternalShearLayers2023}.


\subsection{Inviscid framework in global and local coordinate systems}
In this part, we are going to summarize the general inviscid framework in global and local coordinate systems that will be frequently adopted when developing theories for different configurations.

\begin{figure}
    \centering
    \includegraphics[width=0.7\textwidth]{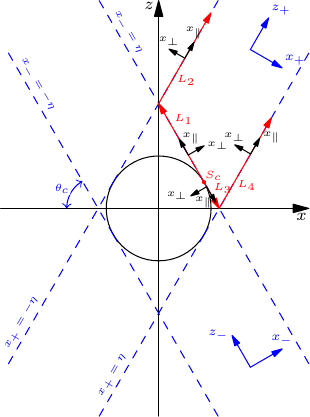}
    \caption{Tangent characteristic lines (blue dashed lines) of a cylinder in an unbounded domain. $(x_\pm, z_\pm)$ are characteristic coordinates. $(x_\parallel, x_\perp)$ are local coordinates attached to critical rays $L_1 (x_-=\eta; z_->0)$, $L_2$ ($x_+=-\eta$; $z_+>\eta$), $L_3$ ($x_-=\eta$; $z_-<0$) and $L_4$ ($x_+=\eta$; $z_+>\eta$) in the first quadrant.}
    \label{fig:unbounded_domain_tangent_characteristics}
\end{figure}

In the inviscid case ($E= 0$), the governing equations (\ref{eq:governing_equations}) can be recast into the famous Poincar\'e equation in the Cartesian coordinate system
\begin{equation}\label{eq:poincare_equation}
    \frac{\partial^2\psi}{\partial x^2}+\left(1-\frac{4}{\omega^2}\right)\frac{\partial^2\psi}{\partial z^2}=0,
\end{equation}
with the streamfunction $\psi$ defined as
\refstepcounter{equation}
$$
 v_x=-\frac{\partial\psi}{\partial z}, \quad v_z=\frac{\partial\psi}{\partial x}. \eqno{(\theequation{\mathit{a},\mathit{b}})}
$$
The velocity component $v_y$ perpendicular to the plane $Oxz$ is related to $v_x$ by
\begin{equation}
    v_y = \frac{2}{\rmi\omega}v_x.
\end{equation}
In the inertial wave regime ($0<\omega<2$), the Poincar\'e equation (\ref{eq:poincare_equation}) is a two-dimensional hyperbolic equation, the solution of which can be described in terms of its characteristics.
As in \citet{voisinBoundaryIntegralsOscillating2021}, the global characteristic coordinates are defined as
\refstepcounter{equation}\label{eq:characteristic_variable}
$$
x_\pm=\sin\theta_cx\mp\cos\theta_cz, \quad z_\pm=\pm\cos\theta_c x+\sin\theta_c z,
\eqno{(\theequation{\mathit{a},\mathit{b}})}
$$
with
\begin{equation}
    \omega=2\cos\theta_c,
\end{equation}
where $\theta_c$ is the acute angle between the characteristics and the horizontal plane.
The characteristic coordinates $x_{\pm}$ and $z_{\pm}$ are perpendicular and parallel to the characteristics, respectively; particularly, $|x_\pm|=\eta$ denote the positions of the characteristics tangent to the cylinder (see blue symbols in figure \ref{fig:unbounded_domain_tangent_characteristics}).
The canonical form of the Poincar\'e equation (\ref{eq:poincare_equation}) is simply $\p^2\psi/\p x_+ \p x_-=0$; the corresponding solution is separable in terms of $x_\pm$, namely,
\begin{equation}\label{eq:global_inviscid_streamfunction}
    \psi(x_+,x_-)=\psi_+(x_+) + \psi_-(x_-).
\end{equation}
The corresponding velocity vector in the $Oxz$ plane is
\begin{equation}\label{eq:global_inviscid_velocity}
v_{z_+}\boldsymbol{\mathrm{e}}_{z_+}+v_{z_-}\boldsymbol{\mathrm{e}}_{z_-}
\end{equation}
with
\begin{equation}\label{eq:global_inviscid_velocity_streamfunction}
    v_{z_\pm}(x_\pm)=\frac{d\psi_\pm}{dx_\pm},
\end{equation}
where $\boldsymbol{\mathrm{e}}_{z_\pm}$ are unit vectors in the $z_\pm$ directions.
The velocity vector can be projected into the Cartesian coordinates, to yield the velocity components in the latter
\refstepcounter{equation}\label{eq:velocity_transform}
$$
v_x=(v_{z_+}-v_{z_-})\cos\theta_c, v_z=(v_{z_+}+v_{z_-})\sin\theta_c, v_y = \frac{2}{\rmi\omega}v_x=-\rmi(v_{z_+}-v_{z_-}).
\eqno{(\theequation{\mathit{a},\mathit{b},\mathit{c}})}
$$

\begin{figure}
    \centering
    \includegraphics[width=0.7\textwidth]{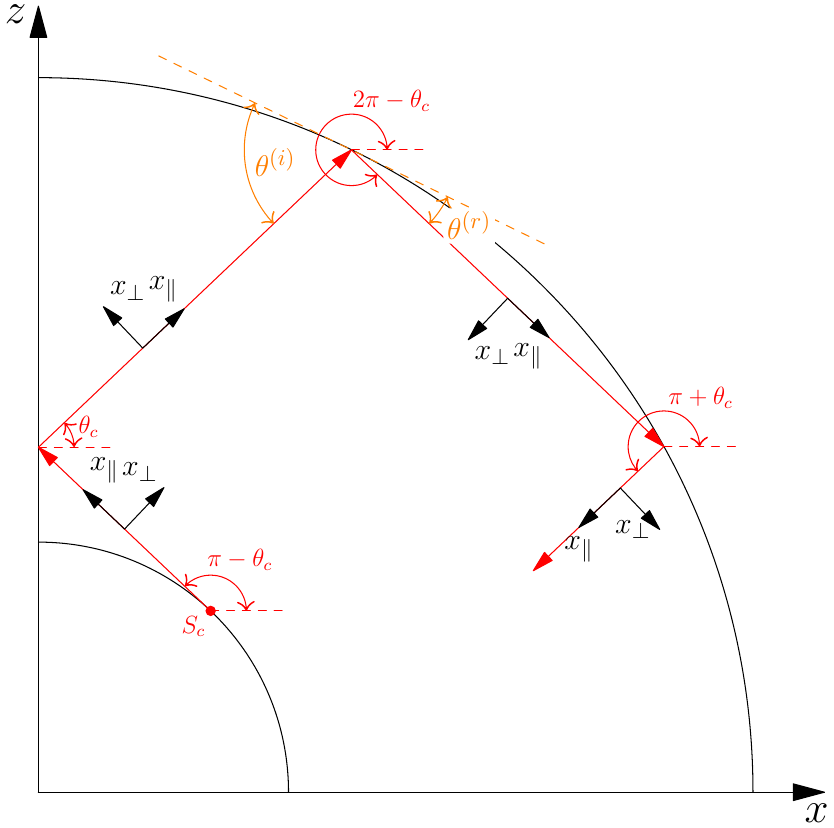}
    \caption{Critical rays with four different propagation directions (in red) and their local coordinates (in black). Schematic of the reflection on a boundary is shown in orange.}
    \label{fig:local_coordinates}
\end{figure}

Usually, the inviscid solution may become singular on critical characteristics.
In this case, it is preferable to describe the solution in terms of local coordinates ($x_\parallel, x_\perp$) around the critical characteristics, where $x_\parallel$ measures the distance to the source along the critical characteristic and $x_\perp$ the displacement relative to it.
The orientations of the local coordinates ($x_\parallel, x_\perp$) depend on the direction of the wave propagation on the critical characteristics.
For a fixed frequency $\omega$ that corresponds to a fixed $\theta_c$, inertial waves propagate in four possible directions, which are $\theta_c$, $\pi-\theta_c$, $\pi+\theta_c$ and $2\pi-\theta_c$ with respect to the positive horizontal direction (see red symbols in figure \ref{fig:local_coordinates}).
The direction of the local parallel coordinate $x_\parallel$ follows the propagation direction; 
the direction of the local perpendicular coordinate $x_\perp$ is chosen such that a nearby non-critical ray remains on the same side (with respect to positive or negative $x_\perp$) of the critical ray after reflection (see black symbols in figure \ref{fig:local_coordinates}).
In these local frames, the streamfunction is related to the local  velocity components $(v_\parallel, v_\perp)$ through 
\refstepcounter{equation}\label{eq:streamfunction_definition_by_local_coordinates}
$$
v_\parallel=\epsilon\frac{\partial\psi}{\partial x_\perp}, \quad v_\perp=-\epsilon\frac{\partial\psi}{\partial x_\parallel},  \eqno{(\theequation{\mathit{a},\mathit{b}})}
$$
where $\epsilon=1$ for the rays with the inclined angles $\pi-\theta_c$ and $2\pi-\theta_c$, and $\epsilon=-1$ for the rays with inclined angles $\theta_c$ and  
$\pi+\theta_c$ 
(see figure \ref{fig:local_coordinates}).
In the inviscid case,  the streamfunction and the parallel velocities depend on the local perpendicular coordinate $x_\perp$ only, namely,
\refstepcounter{equation}
$$
\psi(x_\perp), \quad v_\parallel(x_\perp)=\epsilon\frac{\mathrm{d}\psi(x_\perp)}{\mathrm{d} x_\perp}.  \eqno{(\theequation{\mathit{a},\mathit{b}})}
$$
The velocity components $(v_x, v_z)$ in the Cartesian coordinates are obtained through projecting $v_\parallel\boldsymbol{\mathrm{e}}_\parallel$ into the latter. 
Notably, $v_y$ is related to $v_\parallel$ through a phase shift, namely,
\begin{equation}
    v_{y}=\pm\mathrm{i}v_\parallel.
    \label{eq:vy}
\end{equation}
The sign of the phase is $``+"$ for the rays with the inclined angles $\pi-\theta_c$ and $\pi+
\theta_c$,
and it is $``-"$ for the rays with the inclined angles $\theta_c$ and $2\pi-\theta_c$ (see figure \ref{fig:local_coordinates}).

\subsection{Viscous similarity solution describing internal shear layers}
\label{sec:asymptotic_description}
Now we briefly describe the general asymptotic structure of internal shear layers, irrespective of the  configurations.
We only show the expressions in two dimensions that are relevant to the current work.
Similar descriptions can be found in our previous works \citep{ledizesInternalShearLayers2017,heInternalShearLayers2022,heInternalShearLayers2023}.


In the small viscosity limit ($E\rightarrow 0$), the quantitative feature of the internal shear layers can be described by the famous similarity solution of \citet{mooreStructureFreeVertical1969}, characterised by a width scaling  as $E^{1/3}$.
This similarity solution is expressed in the local coordinates $(x_\parallel, x_\perp)$ described above.
A similarity variable is introduced as
\begin{equation}
    \zeta=\frac{x_{\perp}}{E^{1/3}}\left(\frac{2\sin\theta_c}{x_\parallel}\right)^{1/3}.
\end{equation}
At leading order, the main velocity component is along the critical characteristics, namely $v_\parallel$, which takes the form
\begin{equation}\label{eq:similarity_solution}
v_{\parallel}=C_0H_m(x_\parallel,x_\perp)=C_0\left(\frac{x_\parallel}{2\sin\theta_c}\right)^{-m/3}h_m(\zeta),
\end{equation}
with the special function introduced by \citet{mooreStructureFreeVertical1969}
\begin{equation}\label{eq:moore_saffman_function}
    h_m(\zeta)=\frac{\mathrm{e}^{-\mathrm{i}m\pi/2}}{(m-1)!}\int^{+\infty}_{0}\mathrm{e}^{\mathrm{i}p\zeta-p^3}p^{m-1}dp.
\end{equation}
The perpendicular velocity $v_\perp$ is $O(E^{1/3})$ smaller.
Owing to the fluid rotation, the similarity solution also possesses a velocity component perpendicular to the plane ($\boldsymbol{\mathrm{e}}_{\parallel}$, $\boldsymbol{\mathrm{e}}_\perp$). 
It is related to $v_\parallel$ by a simple phase shift according to the relation (\ref{eq:vy}).

There are two free parameters in the similarity solution (\ref{eq:similarity_solution}): the amplitude $C_0$ and the index $m$.
The calculation of these parameters for each configuration constitutes the core of our work.
It will mainly follow the line of our previous works  \citep{ledizesInternalShearLayers2017,heInternalShearLayers2022,heInternalShearLayers2023} performed 
for a viscous forcing.
One can first note that the similarity solution is indeed associated with an inviscid singularity, which is visible when considering the 
outer limit ($|\zeta| \rightarrow \infty$) of the similarity solution
\begin{equation}\label{eq:outer_limit_of_similarity_solution}
    v_{\parallel}\sim \left\{
    \begin{array}{ll}
      C_0x_\perp^{-m}E^{m/3}, & \zeta\rightarrow +\infty; \\[2pt]
      C_0(-x_\perp)^{-m}\mathrm{e}^{-\mathrm{i}m\pi}E^{m/3},         & \zeta\rightarrow -\infty.
    \end{array}\right.
\end{equation}
Note the behavior in $|x_\perp|^{-m}$ on both sides of the singularity. 
The index $m$ is then directly related to the strength of the underlying inviscid singularity. 
Note also the phase shift in $e^{-{\rm i} m\pi}$ for $x_\perp <0$. This phase shift is associated with the direction of 
propagation of the singular field, which has to be in the direction of $+ {\bf e}_\parallel$   \citep{ledizesCriticalSlopeSingularities2024}.
The similarity solution (\ref{eq:similarity_solution}) can be expressed in terms of a streamfunction $\psi$.
Integration of the similarity solution (\ref{eq:similarity_solution}) leads to
\begin{equation}
    \psi=\epsilon\frac{C_0E^{1/3}}{m-1}H_{m-1}(x_\parallel,x_\perp).
\end{equation}
Note that, the streamfunction $\psi$ is $O(E^{1/3})$ smaller than $v_\parallel$.

\subsubsection{Reflections on boundaries and axes}
In a bounded domain, the self-similar beam is expected to reflect on the boundaries. 
The reflection law, based on the preservation of the self-similar structure and the non-penetrability condition, has been discussed in \citet{ledizesReflectionOscillatingInternal2020,heInternalShearLayers2022,heInternalShearLayers2023}.
For an incident beam $v_{\parallel}^{(i)}=C_0^{(i)}H_m(x_{\parallel}^{(i)},x_{\perp}^{(i)})$ and its reflected counterpart $v_{\parallel}^{(r)}=C_0^{(r)}H_m(x_{\parallel}^{(r)},x_{\perp}^{(r)})$, the conservation of the similarity variable and the non-penetrability condition lead to  relations for the travelled distance and amplitude before and after the reflection
\refstepcounter{equation}\label{eq:reflection_law_on_boundary}
$$
\frac{x_{\parallel b}^{(r)}}{x_{\parallel b}^{(i)}}=\alpha^3, \quad \frac{C_0^{(r)}}{C_0^{(i)}}=\alpha^{m-1},
\eqno{(\theequation{\mathit{a},\mathit{b}})}
$$
where the subscript $b$ indicates that the values are taken at  the reflection point.
The reflection factor $\alpha$ is given by
\begin{equation}
    \alpha=\frac{\sin\theta^{(r)}}{\sin\theta^{(i)}},
\end{equation}
where $\theta^{(r)}$ and $\theta^{(i)}$ represent the angles of the reflected and incident beams relative to the tangent surface at the reflection point (see orange symbols in figure \ref{fig:local_coordinates}).
The value, when compared to $1$, indicates a contraction ($<1$) or expansion ($>1$) of the beam width.
Notably, there is no contraction or expansion on the horizontal or vertical axis, and the reflection factor remains consistently equal to $1$.
The reflection law on a boundary (\ref{eq:reflection_law_on_boundary}) demonstrates that both the distance to the source and the amplitude are a priori modified by reflection.

 The symmetries of the problem with respect to the $Ox$ and $Oz$ axes can be
treated as reflection laws in the reduced domain that we consider (the upper-right quarter). 
Indeed, when a beam reaches the reduced domain boundary corresponding to 
one of the two axes, it effectively crosses the boundary and continues outside the reduced 
domain.
However, the problem symmetry imposes that a symmetrical beam enters the reduced domain as the first beam leaves it.
When this image beam is considered as a reflected beam, we can obtain reflection laws on the two axes as if they were real boundaries.
These laws are obtained by writing the solution close to the axes as 
the sum of these two beams $v_{\parallel}^{(i)}\left(x_{\parallel }^{(i)},x_{\perp }^{(i)}\right)\boldsymbol{\mathrm{e}}_{\parallel}^{(i)}+v_\parallel^{(r)}\left(x_{\parallel }^{(r)},x_{\perp }^{(r)}\right)\boldsymbol{\mathrm{e}}_{\parallel}^{(r)}$. If we apply the conditions of symmetry 
(\ref{eq:symmetry_vx}-\ref{eq:symmetry_vz}) on this expression, we immediately obtain that $v_x$
should vanish on both axes. 
As $x_\perp^{(r)} = x_\perp^{(i)}$ on $Ox$ and  on $Oz$ (see figure \ref{fig:local_coordinates}),
we get that the reflection factor $\alpha$ is 1 for both axes.  
Similar to the reflection law on a boundary (\ref{eq:reflection_law_on_boundary}), we can then write the relations between the two beams as
$$
x_{\parallel b}^{(r)}=x_{\parallel b}^{(i)}, \quad C_0^{(r)}=\rme^{\rmi\varphi}C_0^{(i)},
\eqno{(\theequation{\mathit{a},\mathit{b}})}
$$
where the phase shift $\varphi$ is 
\begin{equation}\label{eq:phase_shift}
    \varphi= \left\{
    \begin{array}{ll}
      \pi, & \quad \mbox{on} \quad Ox, \\[2pt]
      0, & \quad \mbox{on} \quad Oz.
    \end{array}\right.
\end{equation}
Therefore, the ``reflection'' on the $Oz$ axis can be considered as a regular reflection on a vertical surface (i.e. without phase shift) while the ``reflection'' on the $Ox$ axis is a
special reflection on an horizontal surface, as the reflected beam has gained a phase shift of $\pi$.


\section{Unbounded domain}
\label{sec:unbounded_domain}
\begin{figure}
    \centering
    \includegraphics[width=0.5\textwidth]{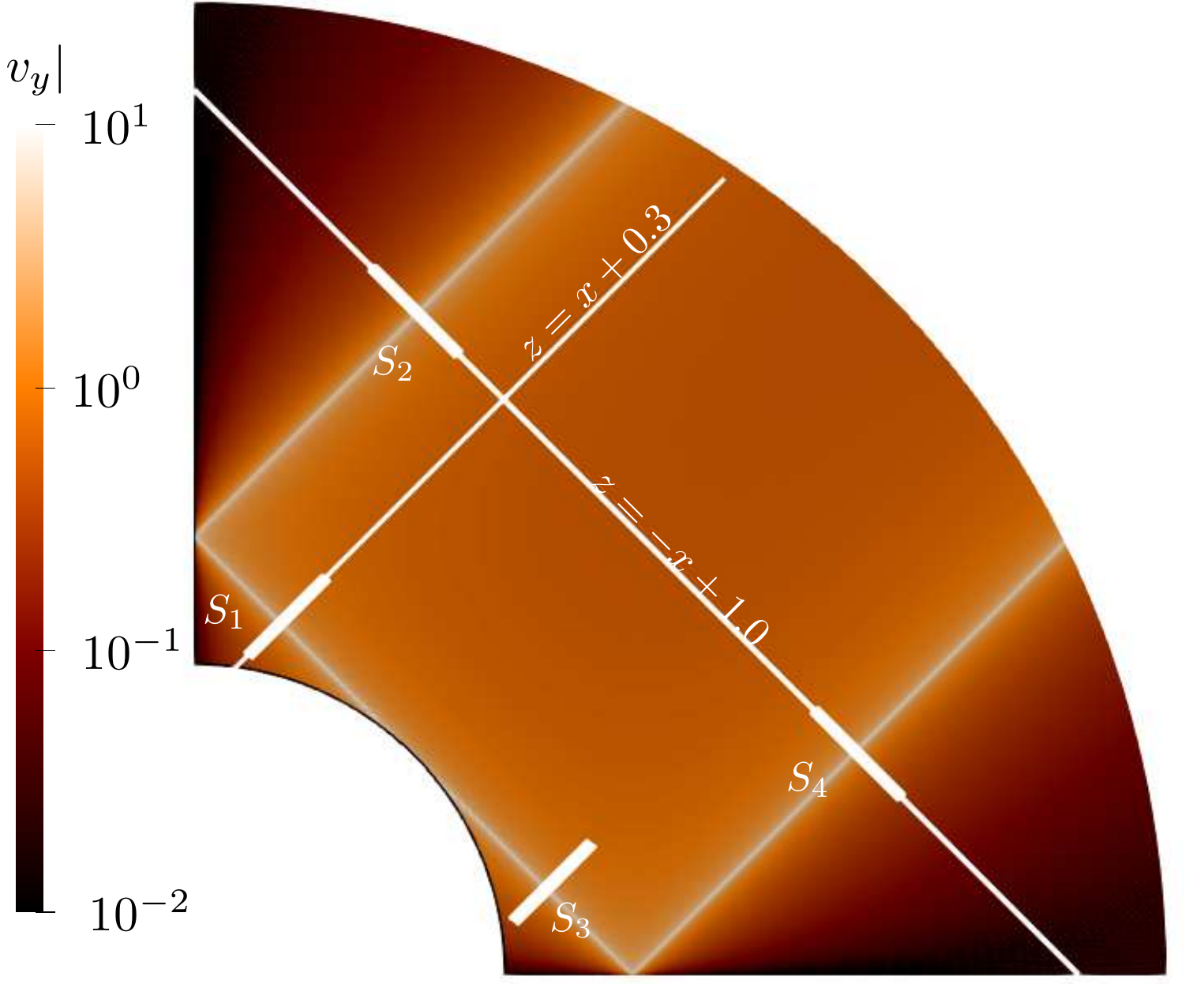}
    \caption{Contour of $|v_y|$ by numerical method for $\omega=\sqrt{2}$, with $\eta=0.35$ and $E=10^{-10}$ in an unbounded domain. Numerical domain size is [0.35, 3.5], with sponge layer specified by $\Upsilon=\omega^{-1}$, $\varrho_s=2.45$ and $\Lambda=0.2$ (see equation \ref{eq:sponge_layer}); the resolution is $N=3000$ and $L=6000$; the result is shown in the truncated domain $[0.35, 1.1]$.}
    \label{fig:unbound_domain_contour}
\end{figure}

We first consider the unbounded domain, without  outer cylinder.
 To illustrate the wave pattern, the numerical contours of $|v_y|$  are shown in figure \ref{fig:unbound_domain_contour} for the frequency $\omega=\sqrt{2}$ at the Ekman number $E=10^{-10}$.
We clearly observe strong shear layers originating from the critical points of the cylinder, where the local slope of the boundary coincides with one of the directions of propagation. 

In order to derive the similarity solutions for the internal shear layers around the tangent characteristics, 
there are two different technical paths.
The first common step is to derive the global inviscid solution.
Afterwards, the two paths diverge.
One path is to obtain the local inviscid solutions around the tangent lines from the global inviscid solution and then obtain the local viscous solutions by matching the local inviscid solutions with the similarity solution.
The other path is to obtain the global viscous solution by adding viscous attenuation to the global inviscid solution and then obtain the local viscous solutions around the tangent lines from the global viscous solution.
One can clearly find that the second method is more advantageous, as no self-similar assumption of the internal shear layers shall be made.
However, we will adopt the first method in the main text, as it is still effective and will also be adopted in the bounded domain.
The details of the second method are shown in Appendix~\ref{app:global_viscous_solution}.
One should note that, the two methods yield the same expressions of the local viscous solutions around the tangent lines (see Appendix~\ref{app:global_viscous_solution}).

\subsection{Global inviscid solution}

The global inviscid solution is obtained by solving the Poincar\'e equation (\ref{eq:poincare_equation}) analytically, 
with the boundary condition at the surface of the cylinder (\ref{eq:boundary_condition}) 
\begin{equation}\label{eq:boundary_condition_on_surface}
    \psi=x \quad \mbox{at} \quad x^2+z^2=\eta^2.
\end{equation}
Note that, we have replaced the no-slip boundary condition with the free-slip counterpart.
At the other end far from the cylinder, the streamfunction is allowed to freely radiate to infinity.

As discussed before, the Poincar\'e  equation (\ref{eq:poincare_equation}) in the inertial range $0<\omega<2$ is actually a hyperbolic equation.
A similar equation, along with the boundary condition (\ref{eq:boundary_condition_on_surface}),  appears in studies of internal wave generation, where the analytical techniques for solving it are well developed \citep{applebyNONBOUSSINESQEFFECTSDIFFRACTION1986,applebyInternalGravityWaves1987,hurleyGenerationInternalWaves1997a,hurleyGenerationInternalWaves1997,voisinLimitStatesInternal2003,voisinInternalWaveGeneration2011,voisinBoundaryIntegralsOscillating2021}.
The general technique involves solving an elliptic equation in the evanescent regime and analytically continuing the solution to the propagating regime.
We can easily adapt this technique to our problem.
We first solve Poincar\'e  equation (\ref{eq:poincare_equation}) in the regime $\omega>2$ where the equation is elliptic.
Using a coordinate stretching similar to that of \citet{voisinBoundaryIntegralsOscillating2021}
\refstepcounter{equation}\label{eq:stretched_cartesian}
$$
x_\star=\frac{\sqrt{\omega^2-4}}{2}x, \quad z_\star=\frac{\omega}{2}z, \eqno{(\theequation{\mathit{a},\mathit{b}})}
$$
the elliptic equation (\ref{eq:poincare_equation}) is recast into a standard Laplacian equation
\begin{equation}
    \frac{\partial^2\psi}{\partial x_\star^2}+\frac{\partial^2\psi}{\partial z_\star^2}=0,
\end{equation}
with the boundary condition specified at the surface of an ellipse
\begin{equation}
    \psi=\frac{2}{\sqrt{\omega^2-4}}x_\star \quad \mbox{at} \quad \left(\frac{2x_\star}{\sqrt{\omega^2-4}\eta}\right)^2+\left(\frac{2z_\star}{\omega\eta}\right)^2=1.
\end{equation}
The Laplacian equation for a region outside an ellipse can be solved analytically using elliptic coordinates
\refstepcounter{equation}\label{eq:elliptic_coordinates}
$$
x_\star=\eta\sinh\sigma\cos\tau, \quad
z_\star=\eta\cosh\sigma\sin\tau, \eqno{(\theequation{\mathit{a},\mathit{b}})}
$$
where $\sigma$ is a non-negative real number and $\tau\in[0,2\pi]$.
Recasting the Laplacian equation into the $(\sigma,\tau)$ coordinates and using separation of variables, we obtain the solution
\begin{equation}\label{eq:unbounded_inviscid_solution_elliptic}
    \psi(\sigma,\tau)=\eta\frac{2}{\omega-\sqrt{\omega^2-4}}\rme^{-\sigma}\cos\tau.
\end{equation}

The next step is to analytically continue the solution (\ref{eq:unbounded_inviscid_solution_elliptic}) from the evanescent regime $\omega>2$ to the propagating regime $0<\omega<2$.
The mathematical details of the analytical continuation are given in  Appendix~\ref{app:analytical_continuation}. 
The solution (\ref{eq:unbounded_inviscid_solution_elliptic}) is analytically continued into the formula (\ref{eq:global_inviscid_streamfunction})
with
\begin{equation}\label{eq:unbounded_domain_inviscid_solution_psi}
    \psi_\pm=\frac{1}{2}\rme^{\rmi(\theta_c-\pi/2)}\left[x_\pm-\sqrt{x_\pm^2-\eta^2}\right],
\end{equation}
where the square root is determined by
\begin{equation}\label{eq:square_root}
  (x_\pm^2-\eta^2)^{1/2} = \left\{
    \begin{array}{ll}
      |x_\pm^2-\eta^2|^{1/2}\sign(x_\pm), & |x_\pm|>\eta; \\[2pt]
      \mp\rmi|x_\pm^2-\eta^2|^{1/2}\sign(z_\pm), & |x_\pm|<\eta.
  \end{array} \right.
\end{equation}
The solution is separable in terms of the characteristic coordinates $x_\pm$, which is a natural property of the two-dimensional hyperbolic equation.

Of particular interest are the velocities along the characteristics.
They can be obtained by taking derivatives of $\psi_{\pm}$ with respect to $x_\pm$ (\ref{eq:global_inviscid_velocity_streamfunction}), which leads to
\begin{equation}\label{eq:unbounded_global_inviscid_solution}
    v_{z_\pm}=\frac{1}{2}\rme^{\rmi(\theta_c-\pi/2)}\left(1-\frac{x_\pm}{\sqrt{x_\pm^2-\eta^2}}\right).
\end{equation}
Thus, we have obtained the global inviscid velocities along the characteristics.
Obviously, the global inviscid solution (\ref{eq:unbounded_global_inviscid_solution}) is singular at the points where $|x_{\pm}|=\eta$, that is on the characteristics tangent to the cylinder. 

\begin{figure}
    \centering
    \includegraphics[width=1.0\linewidth]{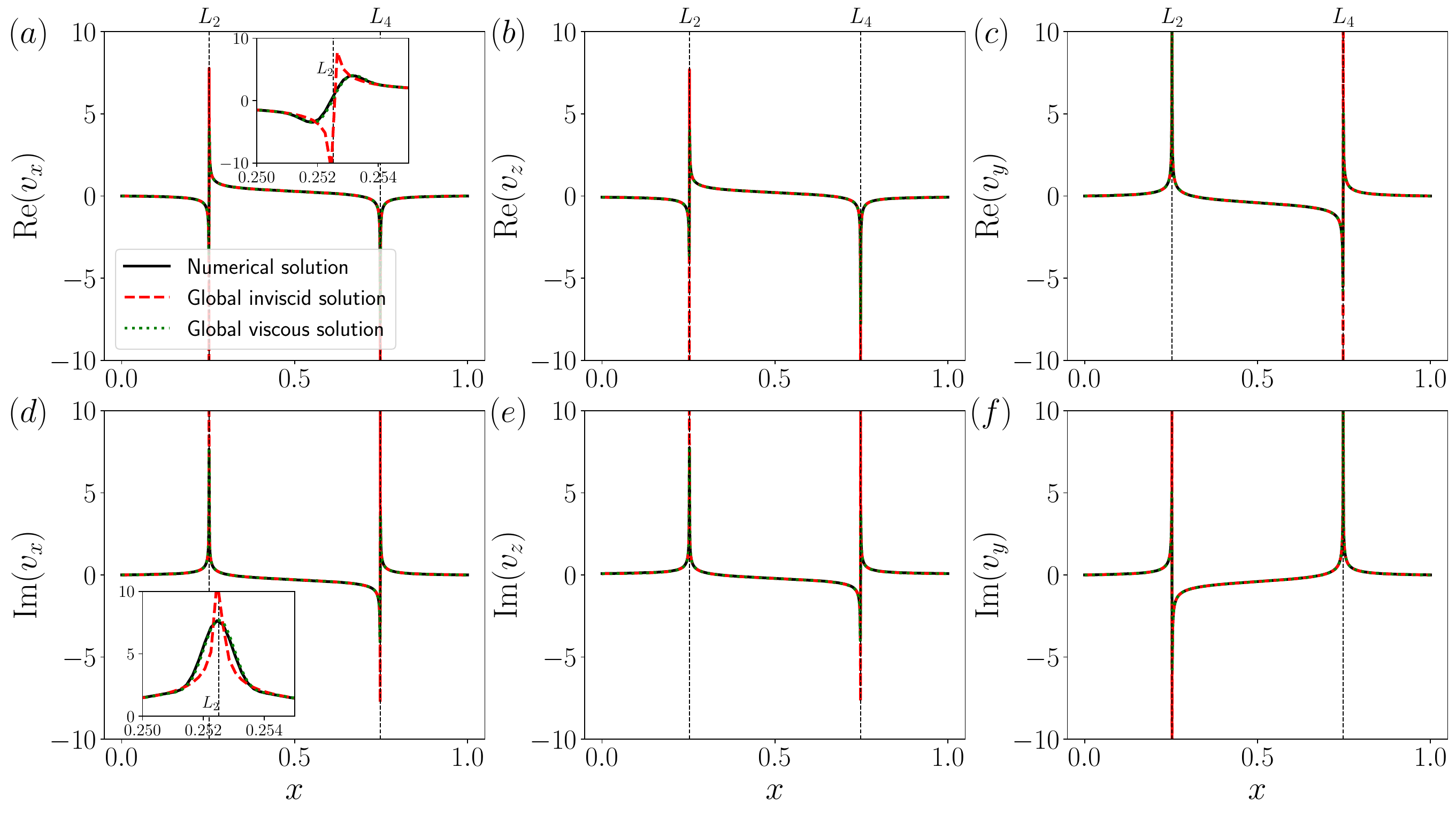}
    \caption{Comparison between numerical, global inviscid (\ref{eq:unbounded_global_inviscid_solution}) and global viscous (\ref{eq:global_viscous_solution_velocity}) solutions for the unbounded case at $E=10^{-10}$ on the line $z=-x+1.0$ (see figure \ref{fig:unbound_domain_contour}) for velocity components $v_x$ ($a,d$), $v_z$ ($b,e$) and $v_y$ ($cf$). 
    Insets show local profiles around the critical line $L_2$. 
    The Jupyter notebook for producing the figure can be found at \url{https://cocalc.com/share/public_paths/c2d25c747f2c92625d0c85d1985e04f0296247d2/figure\%205}.
    }
    \label{fig:unbounded_domain_comparison_z=-x+1.0}
\end{figure}

\begin{figure}
    \centering
    \includegraphics[width=1.0\linewidth]{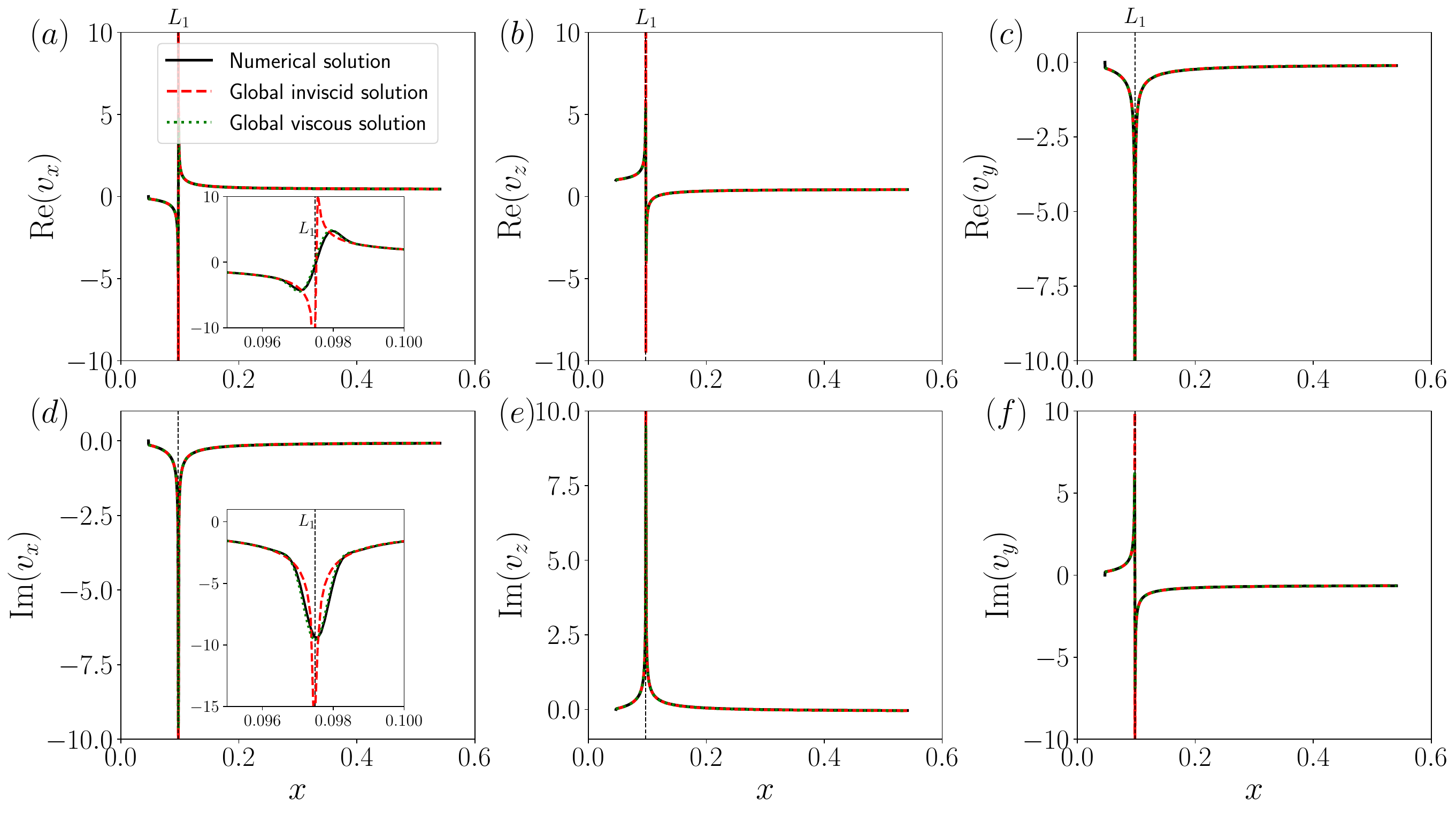}
    \caption{Similar comparison to figure \ref{fig:unbounded_domain_comparison_z=-x+1.0} but for the line $z=x+0.3$ (see figure \ref{fig:unbound_domain_contour}). Insets show local profiles around the critical line $L_1$.
    The Jupyter notebook for producing the figure can be found at \url{https://cocalc.com/share/public_paths/c2d25c747f2c92625d0c85d1985e04f0296247d2/figure\%206}.
    }
    \label{fig:unbounded_domain_comparison_z=x+0.3}
\end{figure}

The velocity components $(v_x, v_z, v_y)$ in the Cartesian coordinates are related to $v_{z_\pm}$ by (\ref{eq:velocity_transform}).
In figures \ref{fig:unbounded_domain_comparison_z=-x+1.0} and \ref{fig:unbounded_domain_comparison_z=x+0.3}, we compare the global inviscid solution (\ref{eq:unbounded_global_inviscid_solution}; red dashed lines) and the numerical counterpart (black solid lines) at $E=10^{-10}$ for the velocity profiles of $(v_x, v_z, v_y)$ along the cuts $z=-x+1.0$ and $z=x+0.3$ (see figure \ref{fig:unbound_domain_contour}), respectively.
The cut $z=-x+1.0$ goes through the critical lines $L_2$ and $L_4$, while $z=x+0.3$ goes through $L_1$ (see figure \ref{fig:unbounded_domain_tangent_characteristics}).
Far from the critical lines, the two solutions agree with each other very well at the small Ekman number.
Close to the critical lines, the singularity of the inviscid solution is clearly visible,  necessitating viscous smoothing.

\subsection{Local inviscid solutions}
Now we try to obtain the local inviscid solutions around the singular lines that will be utilised for asymptotic matching.
Due to symmetry, we only consider the local inviscid solutions near the four critical rays $L_1-L_4$ in the first quadrant; see red arrows in figure \ref{fig:unbounded_domain_tangent_characteristics}.
The definitions of the corresponding local coordinates for the four rays are shown in black in figure \ref{fig:unbounded_domain_tangent_characteristics}.
The transformations between the local and global coordinates perpendicular to the characteristics are as follows
\refstepcounter{equation}
$$
x_{\perp L_1}=x_--\eta; \quad x_{\perp L_2}=-x_+-\eta; \quad x_{\perp L_3}=-x_-+\eta; \quad
    x_{\perp L_4}=-x_++\eta.
\eqno{(\theequation{\mathit{a},\mathit{b},\mathit{c},\mathit{d}})}
$$
The positions of the tangent lines $|x_\pm|=\eta$ become simply $x_\perp=0$ in the local coordinates.
The singular part of the global velocity $v_{z_\pm}$ (\ref{eq:unbounded_global_inviscid_solution}) corresponds to the local parallel velocity $v_\parallel$ close to the critical characteristic.
Thus, we obtain the local inviscid parallel velocity $v_\parallel$ along the lines $L_1-L_4$ as follows
\refstepcounter{equation}\label{eq:unbounded_local_inviscid_solutions}
$$
v_\parallel = \frac{\sqrt{\eta}}{2\sqrt{2}}x_\perp^{-1/2}\left\{
    \begin{array}{ll}
      \rme^{\rmi(\theta_c+\pi/2)}, & L_1; \\[2pt]
      \rme^{\rmi(\theta_c+\pi/2)}, & L_2; \\[2pt]
      \rme^{\rmi\theta_c},         & L_3;  \\[2pt]
      \rme^{\rmi(\theta_c-\pi)}    & L_4.
  \end{array} \right.
  \eqno{(\theequation{\mathit{a},\mathit{b}, \mathit{c}, \mathit{d})}}
$$
The phases are determined by the square roots in the global coordinates according to (\ref{eq:square_root}).
Note that these solutions are only for the positive sides of the local perpendicular coordinates $x_\perp$ (see figure \ref{fig:unbounded_domain_tangent_characteristics}); the counterparts on the negative sides possess an additional phase $\rme^{-\rmi\pi/2}$.

\subsection{Viscous similarity solutions}
The matching of the local inviscid solutions (\ref{eq:unbounded_local_inviscid_solutions}) with the similarity solution (\ref{eq:similarity_solution}) is straightforward.
By doing so, we obtain the index
\begin{equation}
    m=1/2
\end{equation}
and the amplitudes 
\refstepcounter{equation}\label{eq:unbounded_amplitude}
$$
C_0 = \frac{\sqrt{\eta}}{2\sqrt{2}}E^{-1/6}\left\{
    \begin{array}{ll}
      \rme^{\rmi(\theta_c+\pi/2)}, & L_1; \\[2pt]
      \rme^{\rmi(\theta_c+\pi/2)}, & L_2; \\[2pt]
      \rme^{\rmi\theta_c},         & L_3;  \\[2pt]
      \rme^{\rmi(\theta_c-\pi)}    & L_4.
  \end{array} \right.
  \eqno{(\theequation{\mathit{a},\mathit{b}, \mathit{c}, \mathit{d})}}
$$

Obviously, the amplitudes scale with the Ekman number as $E^{-1/6}$.
Note that, there is no phase shift from $L_1$ to $L_2$ on the vertical axis $Oz$, and there is a phase shift of $\pi$ on the horizontal axis $Ox$.
These are consistent with the phase shifts (\ref{eq:phase_shift}) that were obtained from symmetry conditions directly.

\begin{figure}
    \centering
    \includegraphics[width=\textwidth]{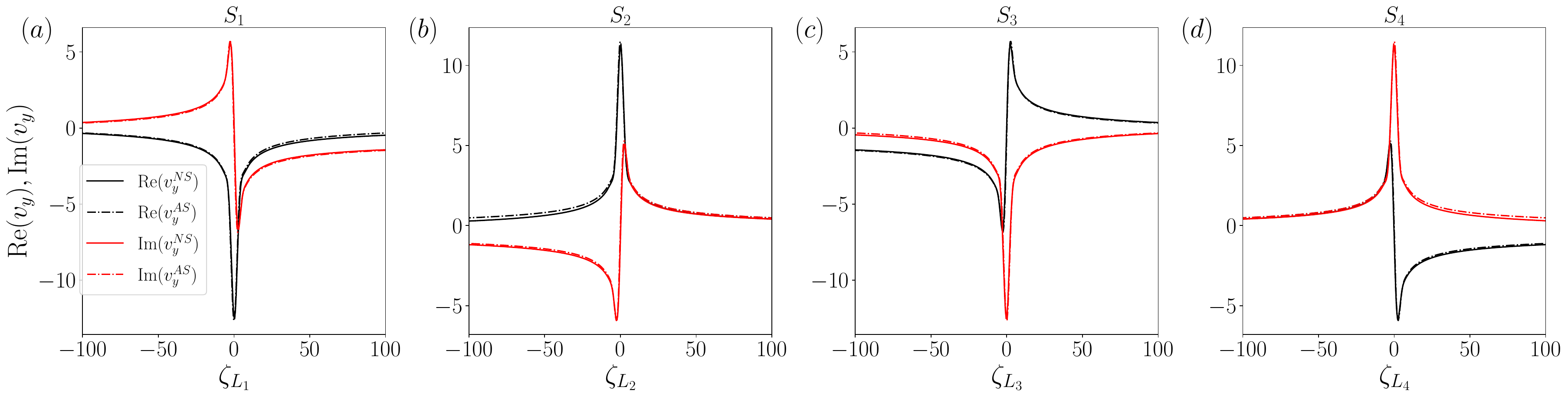}
    \caption{Comparison of velocity profiles between the numerical (superscript ``NS") and asymptotic (superscript ``AS") solutions for $v_y$ at $E=10^{-10}$ on the four short white cuts $S_1-S_4$ shown in figure \ref{fig:unbound_domain_contour}.
    The Jupyter notebook for producing the figure can be found at \url{https://cocalc.com/share/public_paths/c2d25c747f2c92625d0c85d1985e04f0296247d2/figure\%207}.
    }
    \label{fig:open_domain_velocity_profiles}
\end{figure}

After obtaining the index and the amplitudes, we need to determine the local coordinates before calculating the asymptotic solutions.
The coordinate $x_{\perp}$ can be measured according to the definitions shown in figure \ref{fig:unbounded_domain_tangent_characteristics}.
Determining the parallel coordinates $x_\parallel$ requires information about the source position ($x_\parallel=0$).
As in \citet{ledizesInternalShearLayers2017}, the critical point ($S_c$ in figure \ref{fig:unbounded_domain_tangent_characteristics}) is considered as the source, from which two rays propagate northward and southward, traveling along the ray path $L_1\rightarrow L_2$ and $L_3\rightarrow L_4$, respectively.
Thus, $x_\parallel$ for each line is calculated by measuring the distance to the critical point along the ray path.

The excellent performance of the asymptotic solutions against the numerical solutions is shown in figure \ref{fig:open_domain_velocity_profiles} for the velocity profiles $v_y$ at the Ekman number $E=10^{-10}$ around the four critical lines $L_1-L_4$.

\section{Bounded domain}
\label{sec:bounded_domain}
In a bounded domain, the geometry of the characteristics becomes important.
In a cylindrical annulus of fixed aspect ratio, the characteristic patterns only depend on the forcing frequency. As shown by \citet{rieutordInertialWavesRotating2001}, characteristics either form periodic orbits or attractors. 
Both ray patterns were considered in \citet{heInternalShearLayers2022,heInternalShearLayers2023} in the case of a viscous forcing.
In these works, the asymptotic solution was obtained by propagating the similarity solutions from the critical point with the amplitudes obtained from the unbounded domain analysis. 
This approach has been theoretically justified by \citet{ledizesCriticalSlopeSingularities2024}. In that case, the author showed that the characteristics of the similarity solution  only depend on the local property of the viscous forcing close to critical point. 
\citet{ledizesCriticalSlopeSingularities2024} also explained that this result is not valid for an inviscid forcing.
The amplitude of the similarity solution in a closed geometry is expected to depend on the global ray pattern. 

As in the unbounded domain considered in \S~\ref{sec:unbounded_domain}, we shall first derive an inviscid solution.
Unfortunately, a global inviscid solution in a cylindrical annulus cannot be obtained by separation of variables, since the Poincar\'e  equation (\ref{eq:poincare_equation}) is not separable on both boundaries \citep{rieutordInertialWavesRotating2001}.
However, in two dimensions, one can use the method of characteristics.  
The expressions for the streamfunction and velocities (\ref{eq:global_inviscid_streamfunction}-\ref{eq:global_inviscid_velocity_streamfunction}), which are dependent of the global characteristic coordinates $x_\pm$,  are valid in the whole domain. They can be used to obtain equations involving  $v_{z\pm}$  or $\psi_{\pm}$ by propagating the boundary conditions 
along the lines $x_\pm=\text{const}$. This is the method of characteristics \citep{maasGeometricFocusingInternal1995,ogilvieWaveAttractorsAsymptotic2005}
that we shall use for both periodic orbits and attractors.

\subsection{Periodic orbits}
\label{sec:periodic_orbit}
\begin{figure}
    \centering
    \begin{subfigure}{0.55\textwidth}
        \centering
        \includegraphics[width=\textwidth]{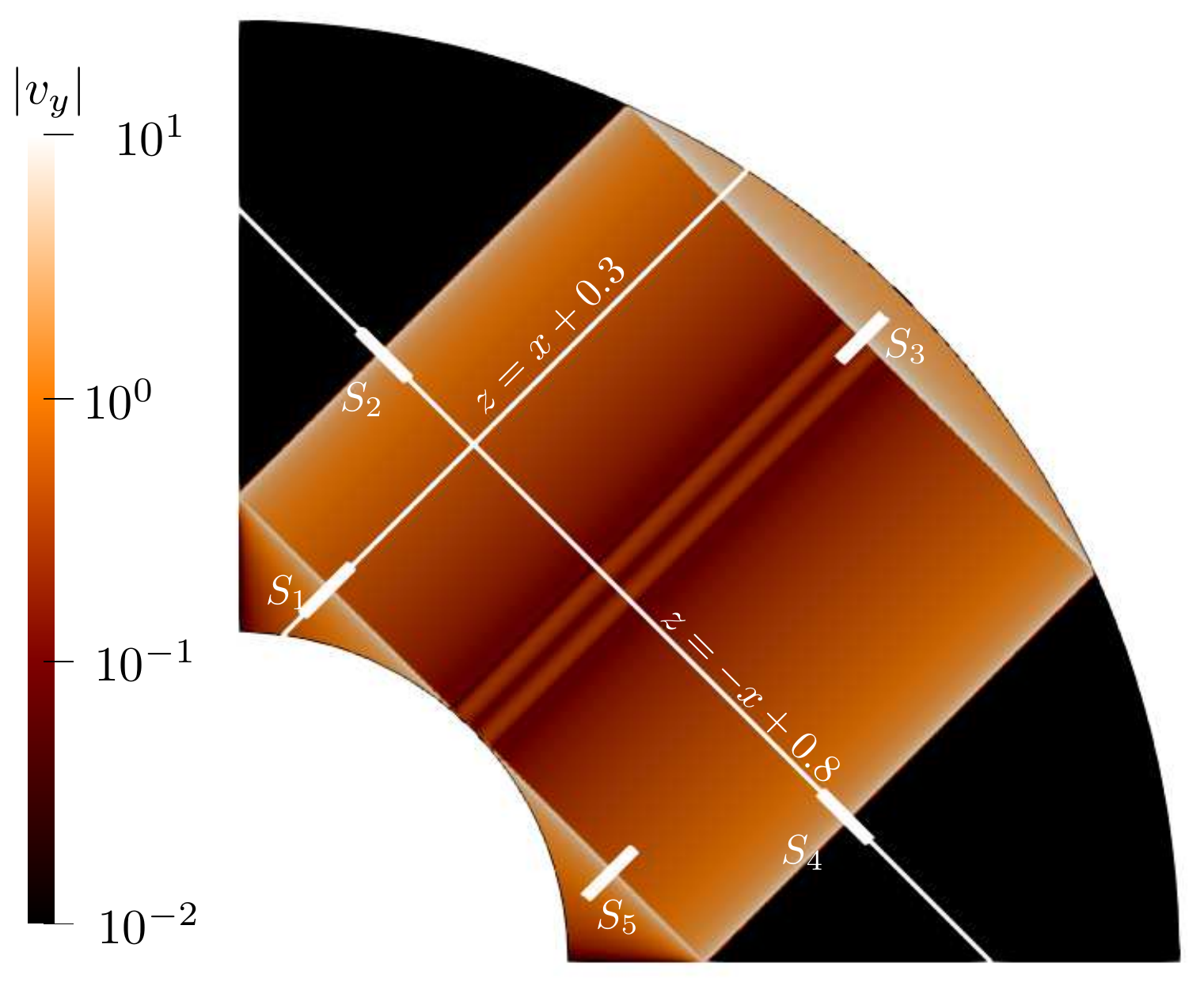}
        \caption{}
    \end{subfigure}%
    \begin{subfigure}{0.45\textwidth}
        \centering
        \includegraphics[width=\textwidth]{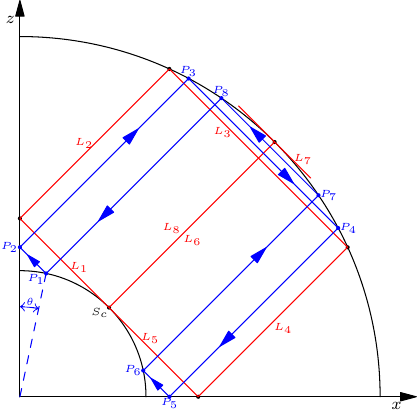}
        \caption{}
    \end{subfigure}
    \caption{Periodic orbit in a bounded domain for $\eta=0.35$ and $\omega=\sqrt{2}$: $(a)$ numerical result of $|v_y|$ at $E=10^{-11}$ with the resolution $(N,L)=(2500,8000)$; $(b)$ critical characteristics $L_1-L_8$ (red color) and a neighboring ray path $P_1P_2P_3P_4P_5P_6P_7P_8$ (blue color) with the co-latitude $\theta$. Note that $L_6$ coincides with $L_8$.}
    \label{fig:periodic_orbit}
\end{figure}

Similar to \citet{heInternalShearLayers2022}, we consider the simple periodic orbit at the frequency $\omega=\sqrt{2}$ and aspect ratio $\eta=0.35$.
The corresponding inclined angle is $\theta_c=\pi/4$.
Figure \ref{fig:periodic_orbit}$(a)$ shows the numerical result for the amplitude of $v_y$ at $E=10^{-11}$, where internal shear layers are observed around the critical characteristics $L_1-L_8$ issued from the critical point $S_c$ (see figure \ref{fig:periodic_orbit}$b$).
One also clearly notices that the solution is much larger  within the area enclosed by the critical lines and  boundaries than in the rest of the domain.

\subsubsection{Global inviscid solutions}
To derive the inviscid solution for this particular ray pattern, we consider an arbitrary circuit issued from a point on the inner boundary different from $S_c$, as shown by the blue lines $P_1P_2P_3P_4P_5P_6P_7P_8P_1$ in figure \ref{fig:periodic_orbit}$(b)$.
The circuit is entirely determined by the co-latitude angle $\theta$ of the point $P_1$ on the inner boundary. 
We denote by $v_{\parallel j}$ the velocity along the segment $P_jP_{j+1}$ ($j=1,2,3,\cdots,8$, $P_9=P_1$) in the direction $P_j\rightarrow P_{j+1}$ as indicated in figure \ref{fig:periodic_orbit}$(b)$. 
If $x_{\pm j}$ denotes the constant value of $x_\pm$ on each segment $P_jP_{j+1}$ using the convention defined in figure \ref{fig:unbounded_domain_tangent_characteristics}, 
the velocity $v_{\parallel j}$  corresponds to  $v_{z_-}(x_{- }) $ for $j= 1,5,7$, $-v_{z_-}(x_{- }) $ for $j=3$, $v_{z_+}(x_{+ }) $ for $j=2,6$ and $-v_{z_+}(x_{+ }) $ for $j=4,8$.

According to the non-penetrability conditions on the boundaries and symmetry conditions on the axes, the velocities $v_{\parallel j}$ of the segments are related to each other by
\begin{subeqnarray}\label{eq:periodic_orbit_boundary_conditions}
    v_{\parallel1}\sin{(\pi/4-\theta)}-v_{\parallel8}\sin{(\pi/4+\theta)}=\cos{\theta}, \\[3pt]
    v_{\parallel2}=v_{\parallel1}, \\[3pt]
    v_{\parallel3}=v_{\parallel2}/k_3, \\[3pt]
    v_{\parallel4}=v_{\parallel3}/k_4, \\[3pt]
    v_{\parallel5}=-v_{\parallel4}, \\[3pt]
    v_{\parallel6}\sin{(\pi/4+\theta)}-v_{\parallel5}\sin{(\pi/4-\theta)}=\sin{\theta}, \\[3pt]
    v_{\parallel7}=v_{\parallel6}/k_7, \\[3pt]
    v_{\parallel8}=v_{\parallel7}/k_8,
\end{subeqnarray}
where the reflection coefficients on the outer boundary are
\refstepcounter{equation}
$$
k_3=\frac{1}{k_4}=\frac{\eta\sin{(\pi/4+\theta)}}{\sqrt{1-\eta^2\sin^2(\pi/4+\theta)}}, \quad k_7=\frac{1}{k_8}=\frac{\eta\sin{(\pi/4-\theta)}}{\sqrt{1-\eta^2\sin^2(\pi/4-\theta)}}. \eqno{(\theequation{\mathit{a},\mathit{b})}}
$$
By symmetry, one has $v_{\parallel 2}=v_{\parallel 4}$, and $v_{\parallel 6} = v_{\parallel 8}$.
The relations (\ref{eq:periodic_orbit_boundary_conditions}) can thus be reduced to two functional equations for $v_{\parallel 1}$ and $v_{\parallel 8}$, namely,
\begin{subeqnarray}\label{eq:periodic_orbit_inviscid_equation}
    v_{\parallel1}\sin{(\pi/4-\theta)}-v_{\parallel8}\sin{(\pi/4+\theta)}=\cos{\theta}, \\[3pt]
    v_{\parallel1}\sin{(\pi/4-\theta)}+v_{\parallel8}\sin{(\pi/4+\theta)}=\sin{\theta}.
\end{subeqnarray}
By solving the above equations for $v_{\parallel 1}$ and $v_{\parallel 8}$ and using the relations for the other components, we can obtain unique expressions for $v_{\parallel j}$ as functions of the co-latitude $\theta$, namely,
\begin{subeqnarray}\label{eq:periodic_orbit_inviscid_solution}
    v_{\parallel1}=v_{\parallel2}=v_{\parallel4}=-v_{\parallel5}=\frac{1}{\sqrt{2}}\frac{\sin{(\pi/4+\theta)}}{\sin{(\pi/4-\theta)}}, \\[3pt]
    v_{\parallel3}=\frac{1}{\sqrt{2}}\frac{\sqrt{1-\eta^2\sin^2(\pi/4+\theta)}}{\eta\sin(\pi/4-\theta)}, \\[3pt]
    v_{\parallel6}=v_{\parallel8}=-\frac{1}{\sqrt{2}}\frac{\sin{(\pi/4-\theta)}}{\sin{(\pi/4+\theta)}}, \\[3pt]
    v_{\parallel7}=-\frac{1}{\sqrt{2}}\frac{\sqrt{1-\eta^2\sin^2(\pi/4-\theta)}}{\eta\sin(\pi/4+\theta)},
\end{subeqnarray}
with $\theta\in[0, \pi/4]$.

The above global inviscid solutions apply only to the area swept by the circuit $P_1\cdots P_8P_1$ with $\theta\in[0, \pi/4]$, which is on one side of the critical lines (see figure \ref{fig:periodic_orbit}($b$)).
On the other side, the ray path does not touch the inner boundary and is not forced.
The solution is therefore undetermined in this region. We choose to put it to zero. 
This choice is consistent with the numerical results displayed in figure \ref{fig:periodic_orbit}($a$): the two regions which are not touched by rays emitted from the inner core are indeed black, that corresponds to a small amplitude. 

The above expressions (\ref{eq:periodic_orbit_inviscid_solution}) can be rewritten in terms of the characteristic variables $(x_+, x_-)$  defined in (\ref{eq:characteristic_variable}).
By doing so, we obtain the functions $v_{z\pm}(x_\pm)$ that appear in (\ref{eq:global_inviscid_velocity}), 
as  we did for the unbounded domain (see expression (\ref{eq:unbounded_global_inviscid_solution})).
The result is
\begin{equation}\label{eq:periodic_orbit_inviscid_vz+}
    v_{z_+}=
\frac{1}{\sqrt{2}}\left\{
    \begin{array}{ll}
      0, & x_+<-\eta; \\
      \frac{-x_+}{\sqrt{\eta^2-x_+^2}}, & -\eta<x_+<\eta; \\
      0,         & x_+>\eta,
  \end{array} \right.
\end{equation}
and
\begin{equation}\label{eq:periodic_orbit_inviscid_vz-}
    v_{z_-}=
    \frac{1}{\sqrt{2}}\left\{
    \begin{array}{ll}
       \sign(x_+)\frac{-x_-}{\sqrt{\eta^2-x_-^2}},  & \frac{\eta}{\sqrt{2}}<x_-<\eta; \\
       0, & \eta< x_- < \sqrt{1-\eta^2}; \\
       \frac{-x_-}{\sqrt{x_-^2-(1-\eta^2)}}, &  \sqrt{1-\eta^2} < x_- < 1.
    \end{array}
    \right.
\end{equation}
Note that  the same term $\sqrt{x_\pm^2-\eta^2}$ appears in formula  (\ref{eq:unbounded_global_inviscid_solution}) and (\ref{eq:periodic_orbit_inviscid_vz+}-\ref{eq:periodic_orbit_inviscid_vz-})  thanks to the use of $\eta$ for the radius of the oscillating cylinder in both the unbounded and bounded  cases.
The expression (\ref{eq:periodic_orbit_inviscid_vz+})  for $v_{z_+}$ is singular at $x_+=\pm\eta$, which correspond to the critical lines $L_2$ and $L_4$;
the expression (\ref{eq:periodic_orbit_inviscid_vz-}) for $v_{z_-}$  is singular at $x_-=\eta$ and $x_-=\sqrt{1-\eta^2}$, which correspond to the critical lines $L_1$, $L_5$ and $L_3$.
For instance, close to $L_1$, one has, as $x_- \rightarrow \eta$
\begin{equation} \label{eq:vz-inviscid-L1}
v_{z_-} \sim \left\{ 
    \begin{array}{ll}
     \displaystyle \frac{\sqrt{\eta}}{2\sqrt{\eta  -x_-}} ,  & x_-<\eta; \\[0.5cm]
       0, & \eta< x_-  .
           \end{array}
    \right.
\end{equation}
Note that on the line $L_6$ (or $L_8$) corresponding to $x_+=0$
, no singularity is present. 

\begin{figure}
    \centering
    \includegraphics[width=1.0\linewidth]{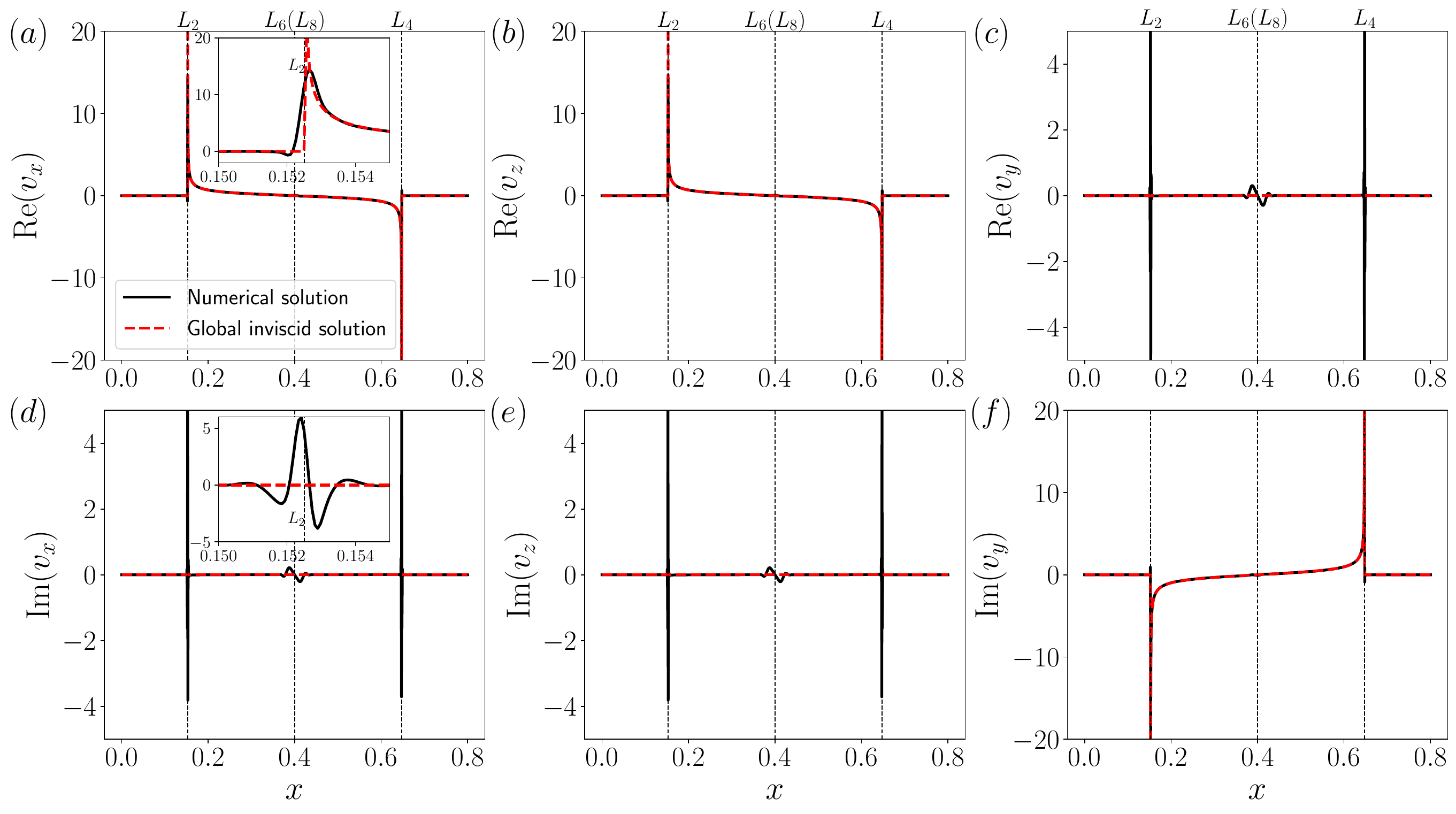}
    \caption{Comparison between inviscid and numerical solutions for the periodic orbit case on the line $z=-x+0.8$ (see figure \ref{fig:periodic_orbit}$a$) for velocity components $v_x$ ($a,d$), $v_z$ ($b,e$) and $v_y$ ($cf$) at $E=10^{-11}$. The vertical axes are in symlog scale. Insets show local profiles around the critical line $L_2$.
    The Jupyter notebook for producing the figure can be found at \url{https://cocalc.com/share/public_paths/c2d25c747f2c92625d0c85d1985e04f0296247d2/figure\%209}.
    }
    \label{fig:periodic_orbit_comparison_on_z=-x+0.8}
\end{figure}

\begin{figure}
    \centering
    \includegraphics[width=1.0\linewidth]{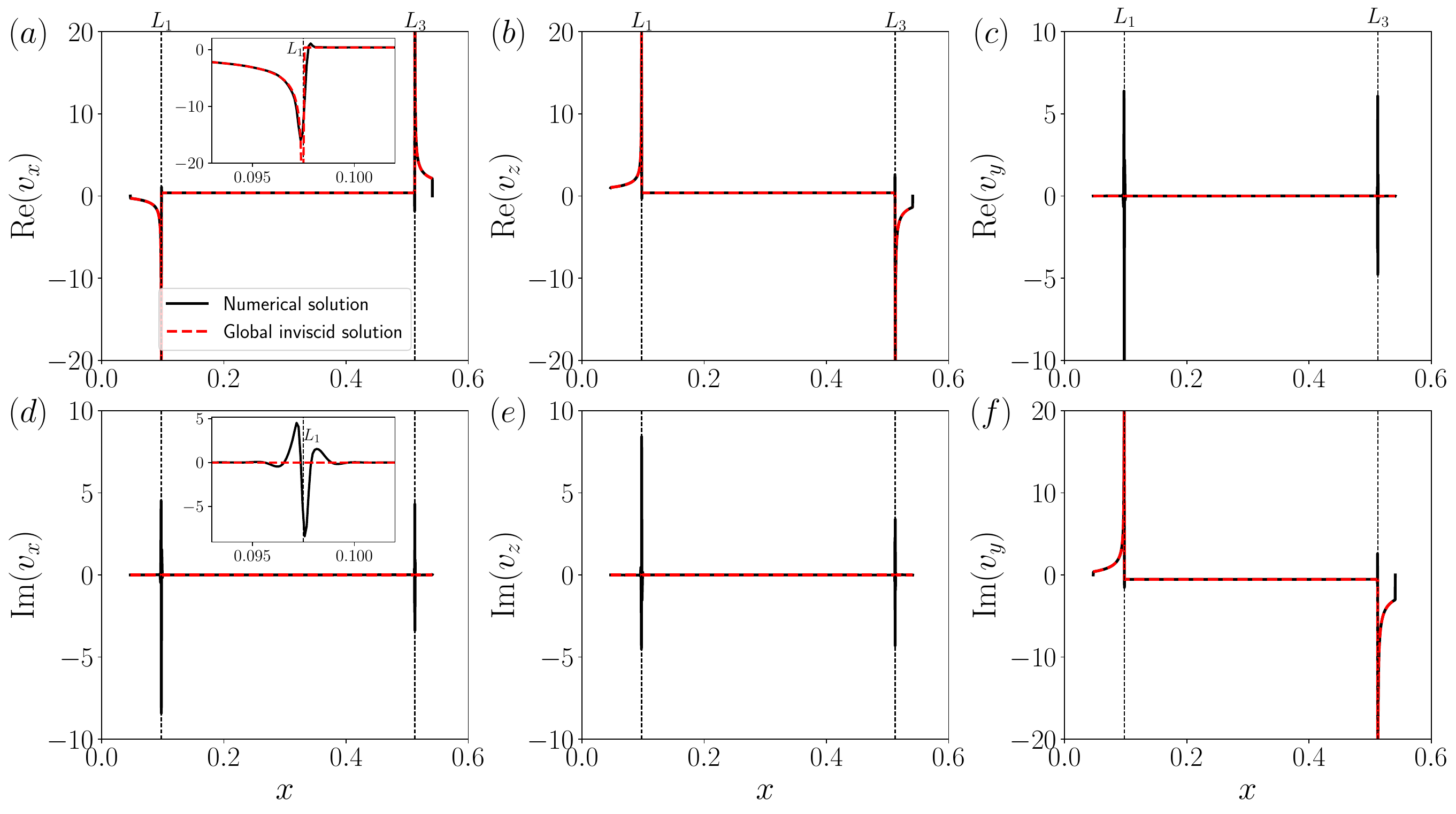}
    \caption{Similar comparison to figure \ref{fig:periodic_orbit_comparison_on_z=-x+0.8} but on the cut $z=x+0.3$ (see figure \ref{fig:periodic_orbit}$a$).
    The Jupyter notebook for producing the figure can be found at \url{https://cocalc.com/share/public_paths/c2d25c747f2c92625d0c85d1985e04f0296247d2/figure\%2010}.
    }
    \label{fig:periodic_orbit_comparison_on_z=x+0.3}
\end{figure}

With the expressions (\ref{eq:periodic_orbit_inviscid_vz+}-\ref{eq:periodic_orbit_inviscid_vz-}) for $v_{z_\pm}$, the velocity components $(v_x, v_z, v_y)$ are obtained through the transforms (\ref{eq:velocity_transform}).
In figures \ref{fig:periodic_orbit_comparison_on_z=-x+0.8} and \ref{fig:periodic_orbit_comparison_on_z=x+0.3}, we compare the global inviscid solution (\ref{eq:periodic_orbit_inviscid_vz+}-\ref{eq:periodic_orbit_inviscid_vz-}) with the numerical solution obtained at $E=10^{-11}$ for the velocity profiles of $(v_x, v_z, v_y)$ along the cuts $z=-x+0.8$ and $z=x+0.3$ (see figure \ref{fig:periodic_orbit}$a$), respectively.
The cut $z=-x+0.8$ goes through the critical lines $L_2$, $L_4$ and $L_6$ (or $L_8$), while $z=x+0.3$ goes through $L_1$ and $L_3$ (see figure \ref{fig:periodic_orbit}$b$).
Far from the critical lines, the two solutions agree with each other very well.
On the other hand, near the critical lines,  the inviscid solution fails to capture the local shear-layer structures (see insets in figures \ref{fig:periodic_orbit_comparison_on_z=-x+0.8}$d$ and \ref{fig:periodic_orbit_comparison_on_z=x+0.3}$d$),
as for the unbounded case. 
Close to the critical lines, viscosity has to be introduced to smooth out the singularity. 
It is the subject of the next subsection.

\subsubsection{Viscous solution close to the critical line}



We are going to construct  a viscous solution close to the critical line following the approach that has been described in \citet{heInternalShearLayers2022}.
The idea is to assume that the critical point $S_c$ is the singularity source, from which two singular beams are generated on either side of $S_c$ and along $L_1$ and $L_5$. These beams are assumed to
have the similarity form (\ref{eq:similarity_solution}) described in section \S \ref{sec:asymptotic_description}.
Based on the nature of the singularity of the inviscid solution in $|x_\perp |^{-1/2}$, one can expect that the similarity solution describing these beams will have an index  $m=1/2$.
Nevertheless, we shall keep  this parameter undetermined as well as 
 the complex amplitudes of the similarity solution generated on each side. 

Starting from the critical point, the northward  beam will travel along the circuit $L_1\rightarrow L_2\rightarrow\cdots \rightarrow L_5$, while the southward beam follows the opposite direction.
$L_1$ and $L_5$ are the initial segments for the northward and southward beams respectively.
Since the ray path is periodic, wave beams will return to the critical point after one cycle. 
They will then continue to propagate for another cycle, and so on, until they are fully dissipated by viscosity.
For the $n$th cycle on the segment $L_j$, the asymptotic solution of the northward or southward beam is denoted as
\begin{equation}
    v_{\parallel j,n}=C_{j,n}H_m(x_{\parallel j,n}, \zeta_{j,n}),
\end{equation}
where the first subscript $j$ denotes the segment $L_j$ and the second subscript $n$ denotes the $n$th cycle ($n=0$ being for the very first cycle).
The expressions of the amplitudes and local coordinates for $n=0$ are determined by reflection laws on the axes and boundaries. 
They have been derived for each segment in \citet{heInternalShearLayers2022}. For completeness, we 
have reproduced these results in appendix~\ref{app:periodic_orbit_formulae}. 
For the subsequent cycles, the perpendicular coordinate remains unchanged, while the amplitude and parallel coordinate are modified, as explained in \citet{heInternalShearLayers2022}.
They can be expressed in terms of the values obtained for the first cycle $n=0$ as
\begin{subeqnarray}
    x_{\parallel {j,n}}=x_{\parallel {j,0}}+n\mathcal{L}_j, \\[3pt]
    \zeta_{j,n} = \zeta_{j,0}(x_{\parallel j,n}/x_{\parallel j,0})^{1/3}, \\[3pt]
    C_{j,n} = C_{j,0} \rme^{\rmi n\pi}.
\end{subeqnarray}
After one cycle, the parallel coordinate increases by the distance traveled along the closed circuit $\mathcal{L}_j$ (see equation \eqref{eq:traveled_distance} below); the amplitude is modified by a phase shift $\pi$ which is acquired at the reflection on the horizontal axis $Ox$.

The final asymptotic solution is the sum of all $v_{\parallel j,n}$ from  $n=0$ to $+\infty$.
As demonstrated by \citet{heInternalShearLayers2022}, this infinite series converges to a finite value if there exists a nonzero phase shift after the completion of each cycle.
This requirement is satisfied as the phase shift along one cycle is $\pi$.
The resulting expression for the segment $L_j$ is 
\begin{equation} \label{exp:vparallel}
    v_{\parallel j} = \lim_{N\rightarrow+\infty}\sum_{n=0}^{N}v_{\parallel j,n}= C_{j,0} G_m(x_{\parallel j,0}, x_{\perp j,0}, \mathcal{L}_j),
\end{equation}
with
\begin{subeqnarray}
    G_m(x_{\parallel},x_{\perp},\mathcal{L})=\left(\frac{x_{\parallel}}{2\sin\theta_c}\right)^{-m/3}g_m(\zeta, \mathcal{L}/x_{\parallel}),\\[3pt]
    g_m(\zeta, \mathcal{L}/x_{\parallel})=\frac{\rme^{-\rmi m\pi/2}}{(m-1)!}\int_0^\infty\frac{\rme^{\rmi p\zeta-p^3}p^{m-1}}{1+\rme^{-p^3\mathcal{L}/x_{\parallel}}}dp.
\end{subeqnarray}
Note that the solution is expressed in terms of the variables in the very first cycle ($n=0$) only.
The above discussion holds both for the northward and southward beams.
Finally, the asymptotic solution 
 is the sum of two contributions, namely,
\begin{equation}\label{eq:periodic_orbit_asymptotic_solution}
    v_{\parallel j}=C_{j,0}^{N} G_m(x_{\parallel j,0}^{N},x_{\perp j,0}^{N},\mathcal{L}_j)+C_{j,0}^{S} G_m(x_{\parallel j,0}^{S},x_{\perp j,0}^{S},\mathcal{L}_j),
\end{equation}
with `N' and `S' denoting `northward' and `southward' respectively. 
The amplitudes and local coordinates for the very first cycle ($n=0$) are given in appendix~\ref{app:periodic_orbit_formulae} in tables \ref{tab:periodic_orbit_northward_beam} and \ref{tab:periodic_orbit_southward_beam} for the northward and southward beams, respectively.
The orientation of $v_{\parallel j}$ is given in figure \ref{fig:local_coordinates}. It corresponds to the orientation of the northward beam. 
From $v_{\parallel j}$, one can deduce the corresponding transverse velocity component $v_y$ using (\ref{eq:vy}). 

We are now able to do the matching of this expression with the behavior of the global inviscid solution close to the critical line. 
The behavior of  the asymptotic solution (\ref{eq:periodic_orbit_asymptotic_solution}) as one goes away from the critical line is 
easily obtained using 
\begin{equation}
    G_m(x_\parallel, x_\perp) \sim \left\{
    \begin{array}{ll}
      \frac{1}{2} x_\perp^{-m}E^{m/3}, & \zeta\rightarrow +\infty \\[2pt]
      \frac{1}{2} (-x_\perp)^{-m}\mathrm{e}^{-\mathrm{i}m\pi}E^{m/3},         & \zeta\rightarrow -\infty.
    \end{array}\right.
\end{equation}
It gives for example on $L_1$ 
\begin{equation}
v_{\parallel 1}=v_{z_-} \sim \left\{
    \begin{array}{ll}
      \frac{1}{2} x_{\perp 1}^{-m}E^{m/3} (C_{1,0}^{N} - \mathrm{e}^{-\mathrm{i}m\pi} C_{5,0}^{S} )   ,  & x_{\perp 1} >0  \\[2pt]
      \frac{1}{2} (-x_{\perp 1})^{-m}E^{m/3}  (C_{1,0}^{N}\mathrm{e}^{-\mathrm{i}m\pi}  -  C_{5,0}^{S} )   ,          & x_{\perp 1} <0 ,
    \end{array}\right.
\end{equation}
where $x_{\perp 1} = x_- -\eta =(x+z-\sqrt{2}\eta)/\sqrt{2}$ is the local coordinate of the northward beam (see table   \ref{tab:periodic_orbit_northward_beam}).
Note that we have replaced $C_{1,0}^{S}$ with $-C_{5,0}^{S}$ using the relation in the second column of the last row of the table \ref{tab:periodic_orbit_southward_beam}.

The matching of these behaviors with the inviscid solution around $L_1$ (equation \eqref{eq:vz-inviscid-L1}) immediately gives $m=1/2$, as expected, and
\refstepcounter{equation}\label{eq:periodic_orbit_amplitude}
$$
C_{1,0}^{N} = \frac{\sqrt{\eta}}{2}\rme^{\rmi\pi/2}E^{-1/6}, \quad C_{5,0}^{S}=\frac{\sqrt{\eta}}{2}E^{-1/6}. \eqno{(\theequation{\mathit{a},\mathit{b})}}
$$
One can then check that with these values, we recover the behavior of the global inviscid solution close to the other lines $L_j$, $j=2,..,5$. 

It is interesting to compare these expressions with those obtained in an unbounded domain (compare (\ref{eq:unbounded_amplitude}$a$) with (\ref{eq:periodic_orbit_amplitude}$a$) for the northward beam and (\ref{eq:unbounded_amplitude}$c$) with (\ref{eq:periodic_orbit_amplitude}$b$) for the southward beam). 
They both scale as $E^{-1/6}$ and exhibit a $\pi/2$ phase shift between northward and southward beams. Yet, the norm and the phase of the beam amplitudes are different in unbounded and bounded cases. 

It is also worth mentioning that (\ref{eq:periodic_orbit_amplitude}) could have been deduced using another argument. 
\citet{ledizesCriticalSlopeSingularities2024} demonstrated that the singular beams emitted along $L_1$ and $L_5$ from $S_c$ depend on the waves emitted along the characteristic direction $L_6$ ($L_8$), see figure \ref{fig:periodic_orbit}(b).
In the unbounded case, the waves emitted in that direction have a specific expression that was deduced from the global inviscid solution.
For the bounded case with periodic orbits, we claim that nothing can be emitted along $L_6$. The reason is the boundary condition on $L_6$ on the outer core which imposes the vanishing of the normal velocity, that is the 
vanishing of the velocity on $L_6$. 
The amplitude of the singular beams emitted along $L_1$ and $L_5$ is then completely determined by the local behavior around $S_c$ in that case. More precisely,
if one  uses equations (49) and (57a,b) of \citet{ledizesCriticalSlopeSingularities2024} with $F'(0)=0$ in (49), expressions (\ref{eq:periodic_orbit_amplitude}a,b) are recovered.


\begin{figure}
    \centering
    \includegraphics[width=\textwidth]{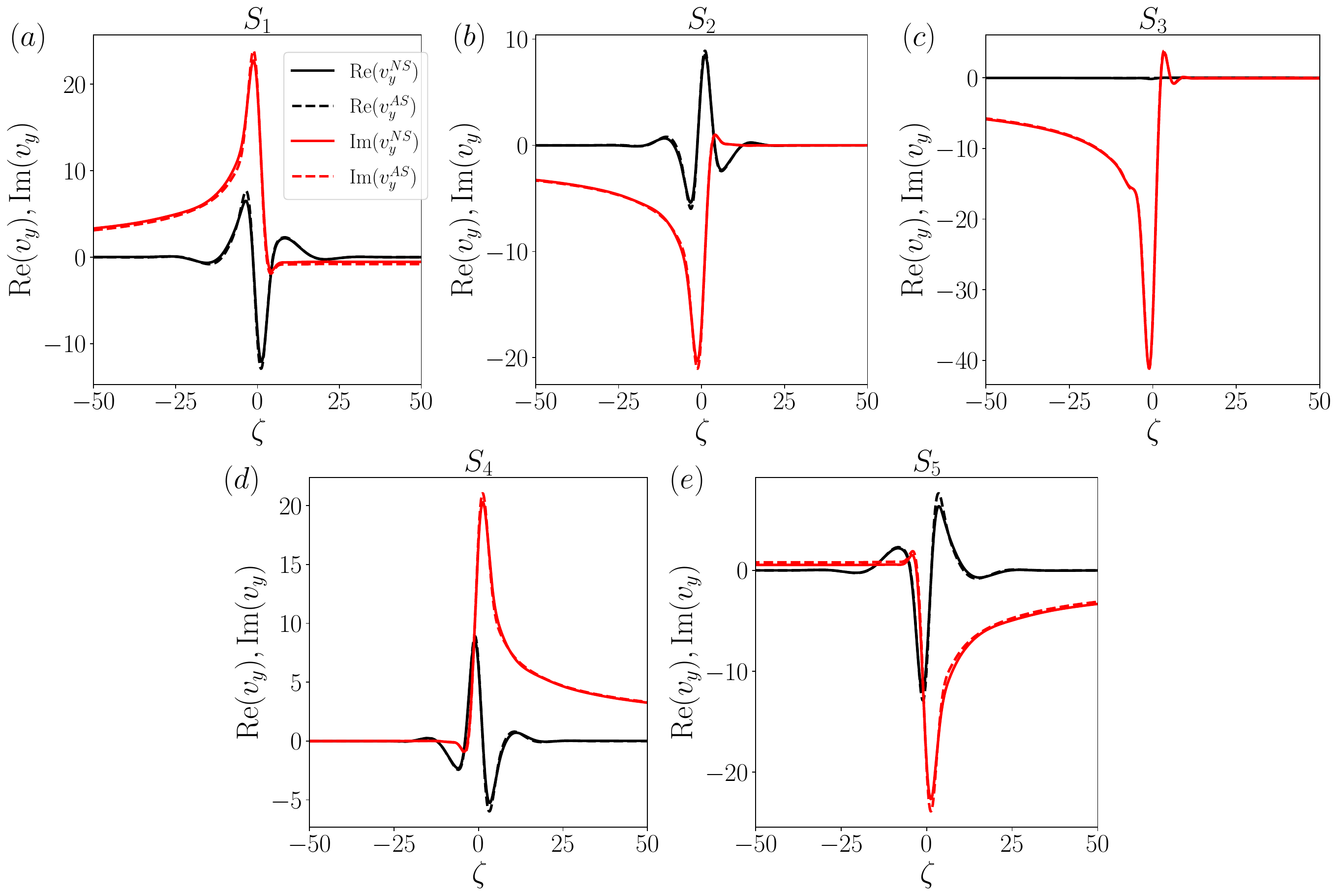}
    \caption{Comparison of velocity profiles ($v_y$) between the numerical and asymptotic solutions for the periodic orbit case at $E=10^{-11}$ on the five cuts $S_1-S_5$ shown in figure \ref{fig:periodic_orbit}.
    The Jupyter notebook for producing the figure can be found at \url{https://cocalc.com/share/public_paths/c2d25c747f2c92625d0c85d1985e04f0296247d2/figure\%2011}.
    }
    \label{fig:periodic_orbit_velocity_profiles}
\end{figure}

In figure \ref{fig:periodic_orbit_velocity_profiles}, the asymptotic solution (\ref{eq:periodic_orbit_asymptotic_solution}) obtained with $C_{j,0}^{N}$ and $C_{j,0}^{S}$ given by (\ref{eq:periodic_orbit_amplitude}$a,b$) is compared with the numerical solution obtained for $E= 10^{-11}$. 
This comparison is done  for the velocity profiles $v_y$ on the five cuts $S_1 - S_5$, shown in figure \ref{fig:periodic_orbit}.
We can see that the agreement is  excellent on all the cuts.  
This confirms that the asymptotic solution derived above describes correctly the solution close to the critical line in the limit of small Ekman numbers.


Note that we have not derived the viscous solutions for the weaker internal shear layers around $L_6$ ($L_8$) (see figure \ref{fig:periodic_orbit} and \ref{fig:periodic_orbit_comparison_on_z=-x+0.8}).
As discussed earlier, they are not related to  inviscid singularities. 
Instead, they result from the split reflections on the inner boundary. 
A related technique for deriving the corresponding viscous solutions, developed by \citet{heInternalShearLayers2022}, can be applied here.
The process is rather straightforward and thus omitted.

\subsection{Attractors}
\label{sec:attractor}
We now consider the more generic case of an attractor. 
The frequency is chosen to be $\omega=0.8102$, the same as in our former work \citep{heInternalShearLayers2023}.
The aspect ratio $\eta=0.35$ is also the same. 
These two parameters are chosen to ensure that the contraction factor of the attractor is significantly different from the two limit values, $1$ and $0$. Additionally, the phase shift along the attractor must be zero to derive an asymptotic solution within our framework.
Numerical results for attractors that do not satisfy the above conditions are shown in Appendix~\ref{app:attractors}.

\begin{figure}
    \centering
    \begin{subfigure}{0.55\textwidth}
        \centering
        \includegraphics[width=\textwidth]{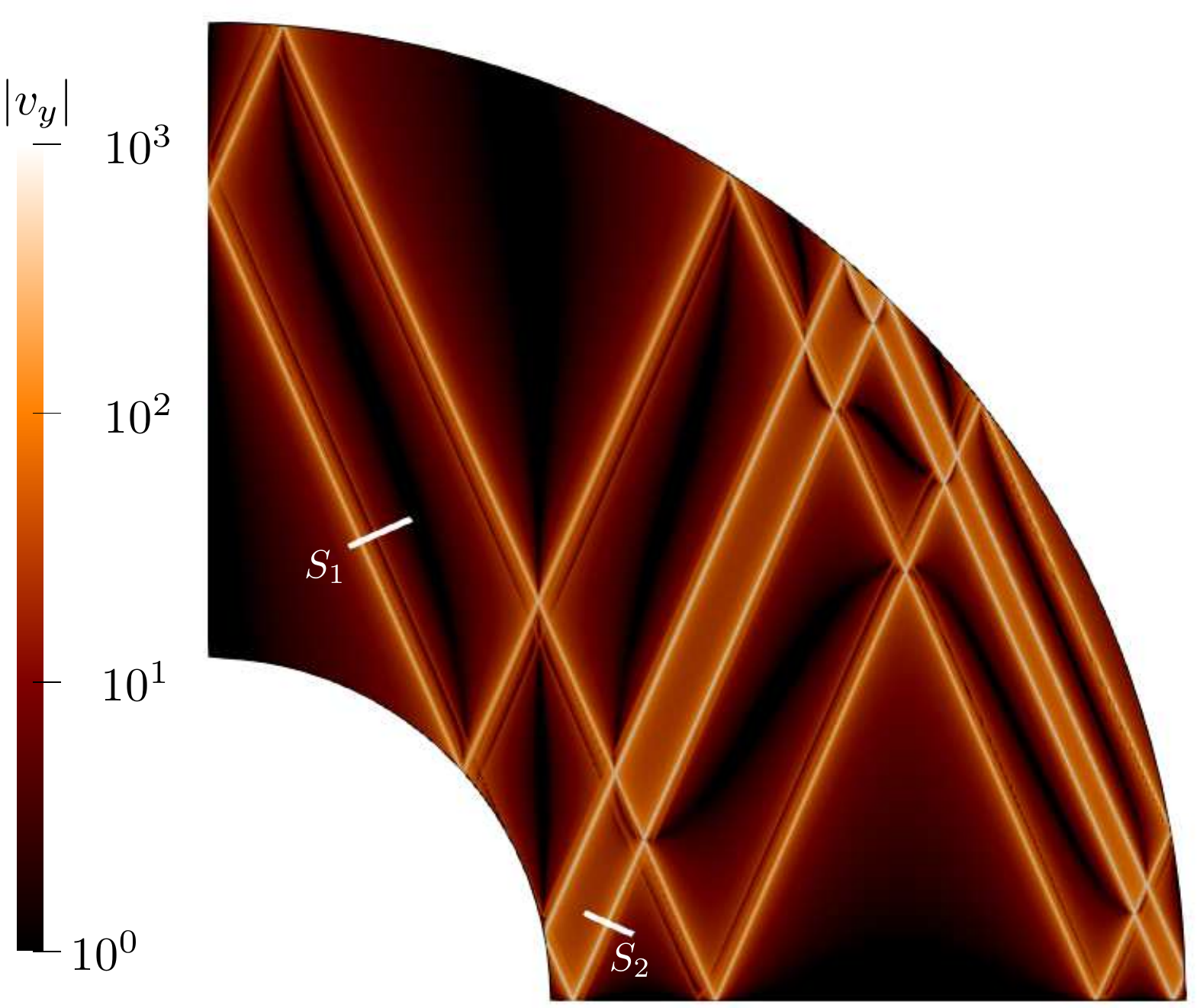}
        \caption{}
    \end{subfigure}%
    \begin{subfigure}{0.45\textwidth}
        \centering
        \includegraphics[width=\textwidth]{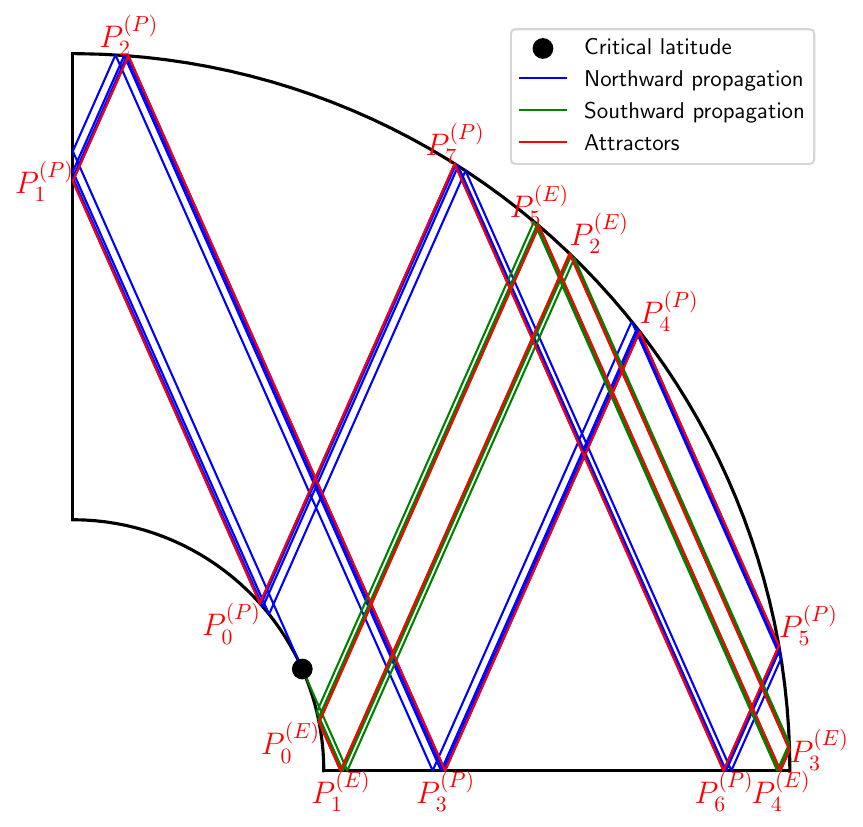}
        \caption{}
    \end{subfigure}
    \caption{Attractors in a bounded domain for $\eta=0.35$ and $\omega=0.8102$. $(a)$ Numerical result of $|v_y|$ at $E=10^{-11}$ with the resolution $(N,L)=(2500,8000)$; $(b)$ Ray paths (blue and green) from the critical point converge to two attractors (red).}
    \label{fig:attractor}
\end{figure}

The qualitative results obtained by numerical method for $E=10^{-11}$ and ray propagation are shown in figure \ref{fig:attractor}.
There are two attractors present: the polar attractor $P_0^{(P)}\cdots P_7^{(P)}$ and the equatorial attractor $P_0^{(E)}\cdots P_5^{(E)}$, reached by the rays propagating northward and southward, respectively.
As discussed by \citet{heInternalShearLayers2023}, when the wave pattern forms attractors, the problem is complicated by the presence of both  critical point and attractor singularities.
For the case of a  viscous forcing,  \citet{heInternalShearLayers2023} have shown that one can construct two asymptotic solutions corresponding to the two underlying singularities, which are valid close to the critical line and attractor, respectively.

For an inviscid forcing, the singular solution generated from the critical point cannot be obtained in a general setting as the amplitude of wave beams emitted along the critical lines is a priori unknown \citep{ledizesCriticalSlopeSingularities2024}. 
However, as we shall see, an asymptotic solution valid around an attractor can be constructed using the method of \citet{ogilvieWaveAttractorsAsymptotic2005}, provided that the attractor has no phase shift \citep{heInternalShearLayers2023}.
This phase shift condition  is satisfied by both attractors shown in figure \ref{fig:attractor}, as both attractors have two contact points on the horizontal axis $Ox$ and the overall phase shift for one cycle is zero.

We now adapt the method developed in \citet{heInternalShearLayers2023} to build an asymptotic solutions close to the attractors.
The first step is to find a local inviscid solution close to each attractor.
By contrast with the periodic orbit considered in \S~\ref{sec:periodic_orbit},  we shall not need a global inviscid solution. 
For convenience, we denote the $J$ vertices of an attractor as $P_0,P_1, \cdots, P_{J-1}$, as shown in figure \ref{fig:attractor}.
The index $J$ is equivalent to the index $0$, namely, $P_J= P_0$.
In the following, the subscript $j$ denotes the variables related to the vertex $P_j$ or the segment $P_jP_{j+1}$. 
The analysis is done  for a  given attractor, either the polar attractor or the equator attractor. 

In \citet{heInternalShearLayers2023}, the method of \citet{ogilvieWaveAttractorsAsymptotic2005} was adapted to derive a functional equation governing the local inviscid streamfunction $\psi_j$ of the segment $P_jP_{j+1}$.
We can simply replace the forcing term of the functional equation with the formula of the vertical oscillation. 
The inviscid functional equation for  an attractor forced by the vertical oscillation is obtained as
\begin{equation}\label{eq:attractor_functional_equation}
    \psi_j(\alpha x_{\perp j})-\psi_j(x_{\perp j}) =\epsilon_j\delta,
\end{equation}
with 
\begin{equation}
    \delta=x_{0},
\end{equation}
and
\begin{equation}\label{eq:contraction_factor}
    \alpha=\alpha_0\alpha_1\cdots\alpha_{J-1}.
\end{equation}
The parameter $\delta$ is the forcing term resulting from the boundary condition imposed by the vertical oscillation of the inner boundary (\ref{eq:boundary_condition_on_surface}); its value is simply the horizontal coordinate $x_0$ of $P_0$, since only $P_0$ is subject to the vertical oscillation.
The parameter $\alpha$ ($<1$) is the contraction factor of the attractor.
The parameter $\epsilon_j$ is the sign in the streamfunction definition (\ref{eq:streamfunction_definition_by_local_coordinates}) in terms of the local coordinates.
The above functional equation means that, after one cycle along the attractor, the beam width is contracted by a factor $\alpha$ while being forced by a constant term $\epsilon_j \delta$ coming 
from the vertical oscillation of the inner boundary. 

The dominant solution to the functional equation (\ref{eq:attractor_functional_equation}) is 
\begin{equation} \label{exp:psi-attractor}
    \psi_j(x_{\perp j}) \sim \frac{\epsilon_j\delta}{\ln \alpha}\ln |x_{\perp j}|.
\end{equation}
According to (\ref{eq:streamfunction_definition_by_local_coordinates}$a$), the corresponding parallel velocity is 
\begin{equation}
    v_{\parallel j}\sim \frac{\delta}{\ln\alpha}x_{\perp j}^{-1},
\end{equation}
by which one can also deduce the transverse velocity $v_{y j}$ using (\ref{eq:vy}).
Once again, we match the above inviscid solution with the outer limit of similarity solution (\ref{eq:outer_limit_of_similarity_solution}).
Finally we obtain the singularity strength 
\begin{equation}
    m=1
\end{equation}
and the amplitude
\begin{equation}\label{eq:attractor_amplitude}
    C_0=\frac{\delta}{\ln\alpha}E^{-1/3}.
\end{equation}
Note that, the contraction factor $\alpha$ should be significantly different from $1$ and $0$ where $\ln^{-1}\alpha$ is singular.
The contraction factors for the polar and equatorial attractors shown in figure \ref{fig:attractor} are $0.354524$ and $0.392994$ respectively, satisfying the above condition.

We need to determine the local coordinates in order to plot the similarity solution.
The perpendicular coordinate $x_{\perp j}$ can be easily calculated by measuring the distance relative to each segment $P_{j}P_{j+1}$.
The parallel coordinate $x_{\parallel j}$ is obtained by an argument which has been detailed in \citet{heInternalShearLayers2023}. 
This coordinate is decomposed into 
\begin{equation}
    x_{\parallel j}= L_j^{(s)}+x_{\parallel j}^\prime,
    \label{exp:xpara}
\end{equation}
where $L_j^{(s)}$ is the distance between $P_j$ and the virtual source,
and $x_{\parallel j}^\prime$, the distance measured from $P_j$ along the segment $P_jP_{j+1}$.
According to the reflection law (\ref{eq:reflection_law_on_boundary}$a$), $L_j^{(s)}$ ($j>0$) is related to the distance  $L_0^{(s)}$ of the first point $P_0$ to the source  by
\begin{equation}
    L_j^{(s)}=(\cdots((L_0^{(s)}+l_0)\alpha_1^3+l_1)\alpha_2^3+\cdots+l_{j-1})\alpha_j^3,
\label{exp:Lj(s)}
\end{equation}
where $\cdots$ stands for a series of left brackets and $l_j$ stands for the length of the segment $P_jP_{j+1}$.
When $j=J$,  the distance $L_J^{(s)}$ is taken to be the same as $L_0^{(s)}$, ensuring that the distance to the virtual source does not change after one complete cycle along the attractor.
This gives an equation for $L_0^{(s)}$ that can be solved as
\begin{equation}\label{eq:attractor_first_source}
    L_0^{(s)}=\left(l_0+\frac{l_1}{\alpha_1^3}+\frac{l_2}{\alpha_1^3\alpha_2^3}+\cdots+\frac{l_{J-1}}{\alpha_1^3\cdots\alpha_{J-1}^3}\right)\frac{\alpha^3}{1-\alpha^3}.
\end{equation}
Using (\ref{exp:xpara}) and (\ref{exp:Lj(s)}), we have then a formula for each parallel coordinate $x_{\parallel j}$. We
are now able to plot the similarity solution for each segment of the attractor.

\begin{figure}
    \centering
    \includegraphics[width=\textwidth]{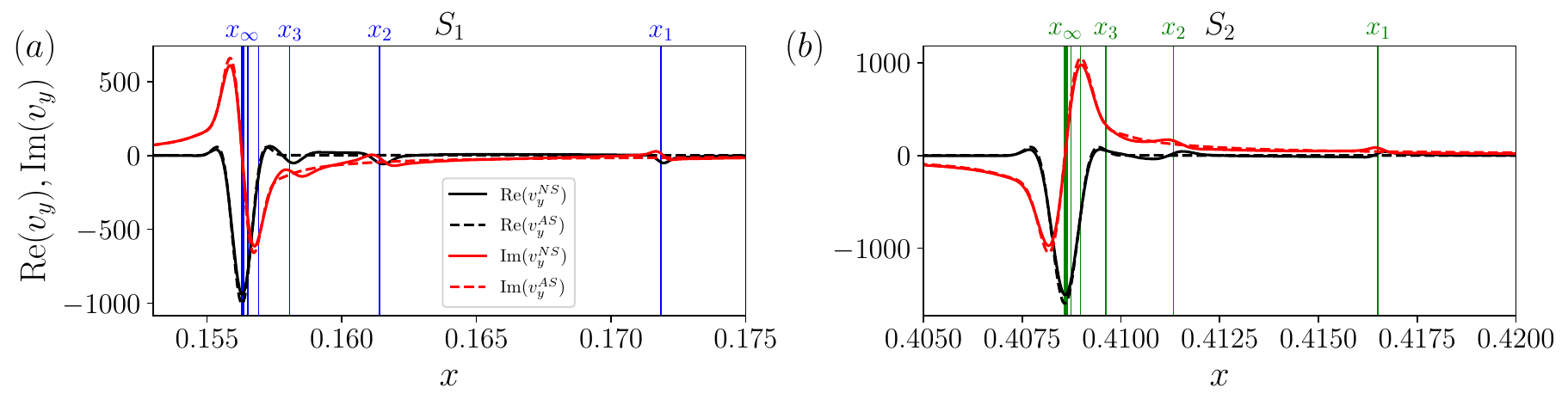}
    \caption{Comparison of velocity profiles ($v_y$) between the numerical and asymptotic solutions for the attractor case at $E=10^{-11}$ on the two cuts shown in figure \ref{fig:attractor}$a$. Blue and green vertical lines are positions of northward and southward critical rays respectively.
    The Jupyter notebook for producing the figure can be found at \url{https://cocalc.com/share/public_paths/c2d25c747f2c92625d0c85d1985e04f0296247d2/figure\%2013}.
    }
    \label{fig:attractor_velocity_profiles}
\end{figure}

The performance of this asymptotic solution against the numerical solution is shown in figure \ref{fig:attractor_velocity_profiles} for the velocity profiles of $v_y$ on the two cuts shown in figure \ref{fig:attractor} at $E=10^{-11}$.
Excellence agreement is achieved around the positions of the attractors ($x_\infty$).
Additionally, in the numerical results, we observe that the numerical solution close to the critical line (that is close to $x_1$, $x_2$, ...) is negligible compared to that around the attractor $x_\infty$.
On the contrary, in the viscous case, the solutions at the two positions are comparable (see figure $8$ of \citet{heInternalShearLayers2023}).

\begin{figure}
    \centering
    \includegraphics[width=0.8\textwidth]{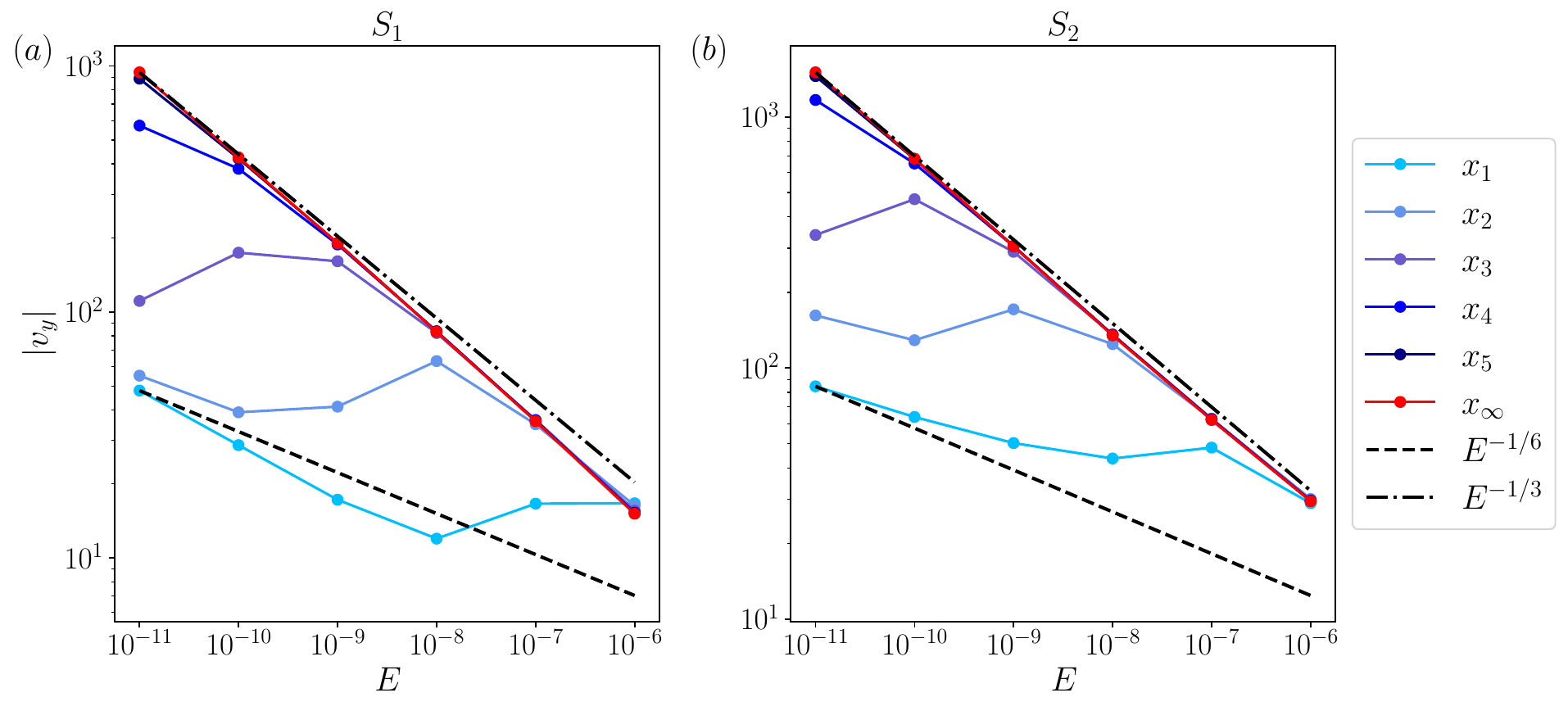}
    \caption{Ekman number scalings of the velocity amplitude at the critical positions on the two cuts shown in figure \ref{fig:attractor}$a$.
    The Jupyter notebook for producing the figure can be found at \url{https://cocalc.com/share/public_paths/c2d25c747f2c92625d0c85d1985e04f0296247d2/figure\%2014}.
    }
    \label{fig:attractor_scaling}
\end{figure}

The Ekman number scalings of the velocity amplitudes at the critical positions on the two cuts are shown in figure \ref{fig:attractor_scaling}.
At the attractor location $x_\infty$, the scaling is $E^{-1/3}$, as predicted.
At the location $x_1$, which is the first crossing point with the critical line, the scaling is close to $E^{-1/6}$, the same as that for the periodic orbit.
This is in agreement with the prediction obtained by \citet{ledizesCriticalSlopeSingularities2024} for the amplitude of the viscous beam created from a critical point by an inviscid forcing.


\section{Conclusion}
\label{sec:conclusion}
Using asymptotic analysis and numerical integration of the linearized governing equations, we have studied the internal shear layers generated in a rotating fluid by vertical oscillations of a two-dimensional cylinder, both in an unbounded and in a bounded domain where the outer boundary is a fixed cylinder with the same axis.
We have shown that these internal shear layers can always be described by the class of similarity solutions introduced by  \citet{mooreStructureFreeVertical1969}.  Our main work has been to find the index (or singularity strength) $m$
and amplitude $C_0$ characterizing these solutions for the different configurations that we have considered.  
This has been done by various techniques. In the unbounded case and in the bounded case with periodic orbits, we have first obtained a global inviscid solution, then performed the asymptotic matching of this solution 
with the local viscous approximation deduced by propagating the similarity solution from the critical point. 
In both cases, we have compared the global inviscid solution and the local viscous approximation with numerical results obtained for small Ekman numbers, and 
an excellent agreement has been observed.
In the bounded case with attractor, only a local inviscid approximation close to the attractor has been derived from which the adequate viscous approximation describing the internal shear layer has been obtained. 
This approximation has  been shown to capture the behaviour of the numerical solution close to the attractor.

In all cases, the width of the internal shear layer is $O(E^{1/3})$, as expected from  the scaling of the similarity solution.  
However, different Ekman scalings for the amplitude $C_0$ have been obtained, with a scaling in $E^{-1/6}$ for the internal shear layer created from a critical point, and in $E^{-1/3}$ for the internal shear layer close to the attractor. 
These large amplitudes are specific to the inviscid forcing. For a viscous forcing, weaker amplitudes in $E^{1/12}$ and $E^{1/6}$ were obtained for the internal shear layers near a critical line and an attractor, respectively 
\citep{ledizesInternalShearLayers2017,heInternalShearLayers2022,heInternalShearLayers2023}. 

These differences in scalings  can be directly attributed to the different index $m$ obtained for the similarity solution describing the internal shear layer, and to the strength of the forcing,
as  $C_0$ is directly obtained by the relation 
\begin{equation}
    C_0\sim O(A_F E^{-m/3}),
\end{equation}
where $A_F$ denotes the forcing magnitude which is $O(E^{1/2})$ for a viscous forcing and $O(1)$ for an inviscid forcing. 
For the inviscid forcing, we have shown that $m=1/2$ for the critical point singularity and $m=1$ for the attractor, while for the viscous forcing, it was $m=5/4$ for the critical point singularity and $m=1$ for the attractor. 
The change of scaling of $C_0$ around an attractor between inviscid and viscous forcings is therefore only due to the change of the strength of the forcing, as the internal shear layer keeps the same similarity structure with $m=1$. 
This is different for the internal shear layer issued from a critical point. In the viscous case,  \citet{ledizesInternalShearLayers2017} obtained an index $m$ larger than 1, that is a singularity stronger than that of the  attractor, while $m$ is smaller than 1 in the inviscid case. 
This explains the dominance of the attractor over the critical point internal shear layer in the inviscid case, while an opposite situation was observed in the viscous case \citep{heInternalShearLayers2023}. 
\begin{table*}
  \begin{center}
  \begin{tabular}{|l|c|c|c}
  \hline
      \diagbox{Forcing}{Singularity} & Critical point & Attractor \\
        \hline
        Viscous forcing & \makecell{$m=5/4$\\$C_0\sim O(E^{1/12})$} &  \makecell{$m=1$\\$C_0\sim O(E^{1/6})$} \\
        \hline
        Inviscid forcing &  \makecell{$m=1/2$\\$C_0\sim O(E^{-1/6})$} &  \makecell{$m=1$\\$C_0\sim O(E^{-1/3})$} \\
       \hline
  \end{tabular}
  \caption{Different values of singularity strength ($m$) and viscous scalings of amplitudes ($C_0$) for different forcings and inviscid singularities.
  Note that, in a bounded domain, the scalings of the critical-point singularity apply to periodic orbits with a phase shift, while those of the attractor singularity apply to attractors without a phase shift.
  }
  \label{tab:summary_parameters}
  \end{center}
\end{table*}
The different values of $m$ and $C_0$ according to the forcing and the nature of the singularity  are summarized in  table  \ref{tab:summary_parameters}.

It is worth recalling the hypotheses that have been made to derive the approximations from which the above scalings have been deduced. 
We have seen that in a bounded domain, the presence of a phase shift has a crucial importance. 
For the periodic orbit case, we have been able to construct a solution because a phase shift was present. 
When there is no phase shift, both the methods used to derive the global inviscid solution and the local viscous approximations break down:
the functional relations   (\ref{eq:periodic_orbit_inviscid_equation}) have no solution, and the sum that gives (\ref{exp:vparallel}) diverges. 
 In that case, we suspect that a resonance with an inviscid eigenmode could occur. 
Inviscid eigenmodes associated with periodic orbits are indeed known to exist \citep{rieutordInertialWavesRotating2001}
and  a resonance was already observed by \cite{rieutordViscousDissipationTidally2010} for a tidal forcing. 
It would be interesting to see whether this situation is possible with our forcing.



For the attractor, no phase shift should be present after one cycle along the attractor in order to construct the local approximation. The reason 
has been explained in \citet{heInternalShearLayers2023}.  With a phase shift, the solution (\ref{exp:psi-attractor})
in $\ln|x_\perp| $ does not exist anymore.  The solution  close to the attractor with a phase shift is therefore not expected to be associated
with a similarity solution of index $m=1$. Its amplitude is also expected to be smaller.
This is in qualitative agreement with the preliminary results that are
provided in Appendix~\ref{app:attractor_with_phase_shift} for an attractor
with a phase shift. A much weaker amplitude is indeed obtained on the attractor in that case,
but surprisingly the $E^{-1/3}$ scaling seems to be still valid. Developing an asymptotic theory  to explain this observation is one of our future objectives. 


Every attractor exists in a frequency range where the associated Lyapunov number
tends to zero and minus infinity at the two ends \citep{rieutordInertialWavesRotating2001}.
The attractor we have considered is at a frequency far from these two ends.
At the frequency where the Lyapunov number is zero, the corresponding contraction factor (\ref{eq:contraction_factor}) tends to $1$. 
At the same time, the amplitude $C_0$ (\ref{eq:attractor_amplitude}) and the distance $L_{0}^{(s)}$
of the first point to the source diverge, suggesting that a stronger solution may exist.
This is confirmed by the numerical example shown in figure \ref{app-fig:attractor_0.8062} of Appendix~\ref{app:attractor_attraction1} where an amplitude scaling as $E^{-1/2}$ is obtained. 
Such a scaling could be related to a resonance with an attractor eigenmode, as first explained by  \cite{rieutordViscousDissipationTidally2010}.  
At the frequency where the Lyapunov number is $-\infty$, the attractor converges towards the critical point.
The attractor solution and the
solution generated from the critical point are not separated anymore. 
The numerical results  shown in figure \ref{app-fig:attractor_0.8249} of Appendix~\ref{app:attractor_attraction0} demonstrate that the solution has an amplitude scaling close to $E^{-0.286}$ in that case. 
An asymptotic theory explaining this scaling  remains to be developed.

In this work, we have only considered two-dimensional (2D) configurations.
Some of the results are expected  to be  also valid in three-dimensional axisymmetric (3DA) configurations. 
For instance, in 3DA, the characteristics in a meridional plane remain the same as in 2D, so the critical lines and the attractors are located at the same place. 
A similarity solution can also be constructed  to describe the internal shear layers far from the axis in 3DA. It has the same expression as in 2D, rescaled by the square root of the distance to the rotation axis.
It has already been used for the case of a viscous forcing \citep{ledizesInternalShearLayers2017,heInternalShearLayers2022,heInternalShearLayers2023}.  
Yet, there is an important difference that is worth mentioning. In 3DA, the similarity solution gains a $\pi/2$ phase jump as it crosses the axis, whereas there is no such phase jump 
 in 2D. This naturally affects the total phase shift obtained on a periodic orbit in 3DA, which will then in general be different  from the 2D problem. 
  In view of the crucial role played by this phase shift in the determination of the amplitude of the similarity solution, one can therefore expect some important effects, especially for the 
 configurations where the phase shift is non-zero in one case, and zero in the other.


\backsection[Acknowledgements]{J.H. acknowledges the China Scholarship Council for financial support (CSC 202008440260). The Centre de Calcul Intensif d’Aix-Marseille is acknowledged for granting access to its high-performance computing resources.
Some of the computations were performed on the HPC3 Cluster of The Hong Kong University of Science and Technology.
The authors are grateful to Bruno Voisin and two  anonymous reviewers  for their insightful comments.
 We would like to particularly thank Bruno Voisin for providing the global viscous solution, which is 
now given in Appendix~\ref{app:global_viscous_solution}.
}


\backsection[Declaration of interests]{The authors report no conflict of interest.}




\appendix

\section{Mathematical details of analytically continuing the solution (\ref{eq:unbounded_inviscid_solution_elliptic})}
\label{app:analytical_continuation}

In this section, we will show how to analytically continue the solution (\ref{eq:unbounded_inviscid_solution_elliptic}) from the evanescent regime ($\omega>2$) to the propagating regime ($0<\omega<2$).
As in \citet{voisinBoundaryIntegralsOscillating2021}, the continuation is implemented with the Lighthill's radiation condition, by adding to the frequency a small positive imaginary part $\epsilon$ which will tend to zero.
Equivalently, the following replacement will be performed 
\begin{equation}
    \omega \rightarrow \omega+\rmi 0=\lim_{\epsilon\rightarrow0^+}(\omega+\rmi\epsilon).
\end{equation}
Following \citet{voisinBoundaryIntegralsOscillating2021}, we aim to express the stretched elliptic coordinates $(\sigma, \tau)$ in terms of the more comprehensible characteristic coordinates ($x_\pm, z_\pm$) (\ref{eq:characteristic_variable}).
Combining (\ref{eq:stretched_cartesian}) and (\ref{eq:elliptic_coordinates}), we have
\refstepcounter{equation}
$$
\sinh{\sigma}\cos{\tau}=\frac{\sqrt{\omega^2-4}}{2}\frac{x}{\eta}, \quad \cosh{\sigma}\sin{\tau}=\frac{\omega}{2}\frac{z}{\eta}.
\eqno{(\theequation{\mathit{a},\mathit{b})}}
$$
Expanding the above expressions with $\omega=2\cos{\theta_c}+\rmi\epsilon$ ($0<\epsilon/2\ll 1$), we obtain
\refstepcounter{equation}
$$
\sinh{\sigma}\cos{\tau}\sim \rmi\frac{x}{\eta}\left(\sin{\theta_c}-\rmi\frac{\epsilon}{2}\cot{\theta_c}\right), \quad
\cosh{\sigma}\sin{\tau}\sim \frac{z}{\eta}\left(\cos{\theta_c}+\rmi\frac{\epsilon}{2}\right).
\eqno{(\theequation{\mathit{a},\mathit{b})}}
$$
Moreover, using the definitions of the characteristic coordinates (\ref{eq:characteristic_variable}), we get 
\begin{equation}
    \cosh{\left(\sigma\mp\rmi\tau-\rmi\frac{\pi}{2}\right)}\sim \frac{x_\pm}{\eta} \mp \rmi\frac{\epsilon}{2}\frac{1}{\sin\theta_c}z_\pm.
\end{equation}
When $\epsilon$ tends to $0$, we can write 
\begin{equation}
    \sigma\mp\rmi\tau -\rmi\frac{\pi}{2}=\mathrm{arcosh}\left(\frac{x_\pm}{\eta}\right).
\end{equation}
Then, replacing $\rme^{-\sigma\pm\rmi\tau}$ in (\ref{eq:unbounded_inviscid_solution_elliptic}) with the above formula, we obtain the solution \eqref{eq:global_inviscid_streamfunction} and \eqref{eq:unbounded_domain_inviscid_solution_psi}.
Note that the term $2/(\omega-\sqrt{\omega^2-4})$ in (\ref{eq:unbounded_inviscid_solution_elliptic}) simply corresponds to $\rme^{\rmi\theta_c}$.

The determination of the square roots (\ref{eq:square_root}) is derived by the replacement
\begin{equation}
    x_\pm\rightarrow x_\pm \mp \rmi 0 \sign(z_\pm).
\end{equation}

\section{Global viscous solution in the unbounded domain}
\label{app:global_viscous_solution}

This appendix is based on a generous contribution from Bruno Voisin. 

In the unbounded domain, a global viscous solution can be obtained from the global inviscid solution, by following \citet{hurleyGenerationInternalWaves1997a,hurleyGenerationInternalWaves1997}.
The global inviscid solution (\ref{eq:unbounded_domain_inviscid_solution_psi},\ref{eq:unbounded_global_inviscid_solution}) can be re-expressed as a spectral integral, with the assistance of a Bessel function $J_1$.
Then an exponential factor accounting for viscous attenuation can be added into the integral.
Using the inverse Fourier transform 
\begin{equation}
    \int_0^\infty J_1(\kappa)\rmexp(\rmi \kappa x)\frac{d\kappa}{\kappa}=\rmi\left[x-\sqrt{(x+\rmi 0)^2-1}\right],
\end{equation}
taken for example from table 5 of \citet{voisinLimitStatesInternal2003}, we have
\begin{equation}
    \psi_\pm=\pm\frac{\eta}{2}\rme^{\rmi\theta_c}\sign z_\pm\int_0^\infty J_1(\kappa\eta)\rmexp(\mp\rmi \kappa x_\pm \sign z_\pm)\frac{d\kappa}{\kappa}.
\end{equation}
Similarly, writing
\begin{equation}
    \int_0^\infty J_1(\kappa)\rmexp(\rmi \kappa x)d\kappa = 1 - \frac{x}{\sqrt{(x+\rmi 0)^2-1}},
\end{equation}
we have 
\begin{equation}
    v_{z_\pm}=\frac{\eta}{2}\rme^{\rmi(\theta_c-\pi/2)}\int_0^\infty J_1(\kappa\eta)\rmexp(\mp\rmi \kappa x_\pm\sign z_\pm)d\kappa.
\end{equation}
The Bessel function $J_1$ in these integrals represents the spatial spectrum of the cylinder, consistent with the application of the boundary integral method to internal gravity waves by \citet{voisinBoundaryIntegralsOscillating2021}; see the second entry in table $5$ there.
Introducing Cartesian coordinates $(k,m)$ and characteristic coordinates $(k_\pm, m_\pm)$ in wavenumber space, related by
\refstepcounter{equation}\label{eq:characteristic_in_wavenumber_space}
$$
k_\pm=k\sin\theta_c \mp m\cos\theta_c, \quad
m_\pm=\pm k\cos\theta_c + m \sin\theta_c, \eqno{(\theequation{\mathit{a},\mathit{b})}}
$$
the complex exponential may be interpreted as the phase factor of each individual plane wave component, $\rmexp(\rmi k_\pm x_\pm+\rmi m_\pm z_\pm)$ with 
\refstepcounter{equation}
$$
k_\pm=\mp\kappa\sign(z_\pm), \quad m_\pm=0, \eqno{(\theequation{\mathit{a},\mathit{b})}}
$$
consistent with the dispersion relation $m_+m_-=0$.

In the presence of viscosity, the equations of motion (\ref{eq:governing_equations}) yield the wave equation
\begin{equation}
    \left[(\rmi\omega+E\nabla^2)^2\nabla^2+4\frac{\partial^2}{\partial z^2}\right]\psi=0,
\end{equation}
with the dispersion relation
\begin{equation}
    \left[\omega+\rmi E(k^2+m^2)\right]^2(k^2+m^2)=4m^2.
\end{equation}
Using characteristic coordinates in wavenumber space (\ref{eq:characteristic_in_wavenumber_space}), the dispersion relation can be expressed as
\begin{equation}
    m_\pm m_\mp=\rmi E(k_\pm^2+m_\pm^2)^2\cos\theta_c - \frac{E^2}{4}(k_\pm^2+m_\pm^2)^3,
\end{equation}
with
\begin{equation}
    m_\mp = \mp k_\pm\sin(2\theta_c) - m_\pm\cos(2\theta_c).
\end{equation}
For small Ekman number $E\ll 1$, this becomes
\begin{equation}
    m_\pm=\mp \rmi\frac{E}{2\sin\theta_c}k_\pm^3.
\end{equation}
turning the streamfunction into
\begin{equation}
    \psi_\pm=\pm\frac{\eta}{2}\rme^{\rmi\theta_c}\sign z_\pm\int_0^\infty J_1(\kappa\eta)\rmexp(-\beta\kappa^3|z_\pm|)\rmexp(\mp\rmi\kappa x_\pm \sign z_\pm)\frac{d\kappa}{\kappa},
\end{equation}
and the velocity into
\begin{equation}\label{eq:global_viscous_solution_velocity}
    v_{z_\pm}=\frac{\eta}{2}\rme^{\rmi(\theta_c-\pi/2)}\int_0^\infty J_1(\kappa\eta)\rmexp(-\beta\kappa^3|z_\pm|)\rmexp(\mp\rmi\kappa x_\pm\sign z_\pm)d\kappa,
\end{equation}
with
\begin{equation}
    \beta=\frac{E}{2\sin\theta_c}.
\end{equation}
In this way, adapting the approach of \citet{hurleyGenerationInternalWaves1997a} and \citet{hurleyGenerationInternalWaves1997}, a global viscous solution has been obtained.
It is similar, given the different geometry, to the solution of \citet[eqs. (3.21)-(3.22)]{LeDizes2015} and \citet[eq. (3.4)]{ledizesInternalShearLayers2017} for the waves from a librating disk.
\citet{voisinNearfieldInternalWave2020} has shown that this solution describes the wave structure
both in the far field and in the near field, provided the assumption $E\ll 1$ holds. 

In figures \ref{fig:unbounded_domain_comparison_z=-x+1.0} and \ref{fig:unbounded_domain_comparison_z=x+0.3}, we compare the numerical solution (black solid lines) with the global viscous solution (\ref{eq:global_viscous_solution_velocity}; green dotted lines) for $E=10^{-10}$. One can see that they closely agree with each other everywhere.

For small $E$, close to the critical rays, the main contribution to the integral (\ref{eq:global_viscous_solution_velocity}) comes from large wavenumbers. Replacing the 
Bessel function by its expansion for large arguments 
\begin{equation}
    J_1(\kappa\eta)\sim \sqrt{\frac{2}{\pi\kappa\eta}}\sin(\kappa\eta-\pi/4),
\end{equation}
gives
\begin{eqnarray}
v_{z_\pm} &\sim & \frac{\eta^{1/2}}{2^{3/2}}\rme^{\rmi\theta_c}\left(\frac{2\sin\theta_c}{E|z_\pm|}\right)^{1/6} \biggl\{ \rmi h_{1/2}\left[\left(\frac{2\sin\theta_c}{E|z_\pm|}\right)^{1/3}(\mp x_\pm\sign z_\pm-\eta)\right] \nonumber\\
&& -h_{1/2}\left[\left(\frac{2\sin\theta_c}{E|z_\pm|}\right)^{1/3}(\mp x_\pm\sign z_\pm+\eta)\right] \biggr\},
\end{eqnarray}
with $h_{1/2}$ the Moore-Saffman function (\ref{eq:moore_saffman_function}).
For the rays $L_1$ to $L_4$, this gives the local expressions in table \ref{tab:local_expansions}, 
consistent with (\ref{eq:similarity_solution}) and (\ref{eq:unbounded_amplitude}).

\begin{table}
  \begin{center}
\def~{\hphantom{0}}
  \begin{tabular}{lccc}
      Critical line  & $x_\perp$   &   $x_\parallel$ & $v_\parallel$ \\[3pt]
       $L_1$   & $x_--\eta$ & $z_-$ & $\frac{\eta^{1/2}}{2^{3/2}}\rme^{\rmi(\theta_c+\pi/2)}\left(\frac{2\sin\theta_c}{Ex_\parallel}\right)^{1/6}h_{1/2}(\zeta)$ \\[3pt]
       $L_2$   & $-x_+-\eta$ & $z_+$ & $\frac{\eta^{1/2}}{2^{3/2}}\rme^{\rmi(\theta_c+\pi/2)}\left(\frac{2\sin\theta_c}{Ex_\parallel}\right)^{1/6}h_{1/2}(\zeta)$ \\[3pt]
       $L_3$   & $-x_-+\eta$ & $-z_-$ & $\frac{\eta^{1/2}}{2^{3/2}}\rme^{\rmi\theta_c}\left(\frac{2\sin\theta_c}{Ex_\parallel}\right)^{1/6}h_{1/2}(\zeta)$ \\[3pt]
       $L_4$   & $-x_++\eta$ & $z_+$ & $-\frac{\eta^{1/2}}{2^{3/2}}\rme^{\rmi\theta_c}\left(\frac{2\sin\theta_c}{Ex_\parallel}\right)^{1/6}h_{1/2}(\zeta)$ \\
  \end{tabular}
  \caption{Local expansions of the waves close to the critical rays.}
  \label{tab:local_expansions}
  \end{center}
\end{table}

\section{Properties of the similarity solution on the first cycle for the periodic orbit case ($\theta_c=\pi/4$).} 
\label{app:periodic_orbit_formulae}

In this section, we provide the value of the quantities needed to define the similarity solution (\ref{eq:similarity_solution})  on the first cycle of the periodic orbit issued from the critical point $S_c$ 
for the case $\theta_c =\pi/4$. 

The results are shown in tables \ref{tab:periodic_orbit_northward_beam}-\ref{tab:periodic_orbit_southward_beam} for the northward and southward beams, respectively.
The amplitude $C_{j,0}$ and the local coordinates ($x_{\parallel j,0}, x_{\perp j,0}$) for the first cycle are given for each segment $L_1,...,L_5$ defined in figure \ref{fig:periodic_orbit}($b$).
Note that, the amplitudes are expressed in terms of $C_{1,0}^{N}$ and $C_{5,0}^{S}$ for the northward and southward beams respectively.
The expressions of $C_{1,0}^{N}$ and $C_{5,0}^{S}$ are given by (\ref{eq:periodic_orbit_amplitude}).

\begin{table}
  \begin{center}
\def~{\hphantom{0}}
  \begin{tabular}{lccc}
      Critical line  & Amplitude $C_{j,0}^{N}$   &   $x_{\parallel j,0}$ & $x_{\perp j,0}$ \\[3pt]
       $L_1$   & $C_{1,0}^{N}$ & $\frac{-x+z}{\sqrt{2}}$ & $\frac{x+z-\sqrt{2}\eta}{\sqrt{2}}$ \\[3pt]
       $L_2$   & $C_{1,0}^{N}$ & $l_1+\frac{x+z-\sqrt{2}\eta}{\sqrt{2}}$ & $\frac{-x+z-\sqrt{2}\eta}{\sqrt{2}}$ \\[3pt]
       $L_3$   & $C_{1,0}^{N}k_{3c}^{-1/2}$ & $(l_1+l_2)k_{3c}^3+\frac{x-z+\sqrt{2}\eta}{\sqrt{2}}$ & $\frac{-x-z+\sqrt{2-2\eta^2}}{\sqrt{2}}$ \\[3pt]
       $L_4$   & $C_{1,0}^{N}$ & $l_1+l_2+l_3k_{3c}^{-3}+\frac{-x-z+\sqrt{2-2\eta^2}}{\sqrt{2}}$ & $\frac{x-z-\sqrt{2}\eta}{\sqrt{2}}$ \\[3pt]
       $L_5$   & $-C_{1,0}^{N}$ & $l_1+l_2+l_3k_{3c}^{-3}+l_4+\frac{-x+z+\sqrt{2}\eta}{\sqrt{2}}$ & $\frac{x+z-\sqrt{2}\eta}{\sqrt{2}}$ \\
  \end{tabular}
  \caption{Amplitudes and local coordinates of the northward beam from $L_1$ to $L_5$.}
  \label{tab:periodic_orbit_northward_beam}
  \end{center}
\end{table}

\begin{table}
  \begin{center}
\def~{\hphantom{0}}
  \begin{tabular}{lccc}
      Critical line  & Amplitude $C_{j,0}^{S}$   &   $x_{\parallel 5,0}$ & $x_{\perp j,0}$ \\[3pt]
       $L_5$   & $C_{5,0}^{S}$ & $\frac{-x-z+\sqrt{2}\eta}{\sqrt{2}}$ & $\frac{-x-z+\sqrt{2}\eta}{\sqrt{2}}$ \\[3pt]
       $L_4$   & $-C_{5,0}^{S}$ & $l_5+\frac{x+z-\sqrt{2}\eta}{\sqrt{2}}$ & $\frac{-x+z+\sqrt{2}\eta}{\sqrt{2}}$ \\[3pt]
       $L_3$   & $-C_{5,0}^{S}k_{4c}^{1/2}$ & $(l_5+l_4)k_{4c}^{-3}+\frac{-x+z+\sqrt{2}\eta}{\sqrt{2}}$ & $\frac{x+z-\sqrt{2-2\eta^2}}{\sqrt{2}}$ \\[3pt]
       $L_2$   & $-C_{5,0}^{S}$ & $l_5+l_4+l_3k_{4c}^{3}+\frac{-x-z+\sqrt{2-2\eta^2}}{\sqrt{2}}$ & $\frac{x-z+\sqrt{2}\eta}{\sqrt{2}}$ \\[3pt]
       $L_1$   & $-C_{5,0}^{S}$ & $l_5+l_4+l_3k_{4c}^{3}+l_2+\frac{x-z+\sqrt{2}\eta}{\sqrt{2}}$ & $\frac{-x-z+\sqrt{2}\eta}{\sqrt{2}}$ \\
  \end{tabular}
  \caption{Amplitudes and local coordinates of the southward beam from $L_5$ to $L_1$.}
  \label{tab:periodic_orbit_southward_beam}
  \end{center}
\end{table}
The lengths $l_1,..,l_5$ of the segment $L_1,..,L_5$ that appear in the expression of $x_\parallel$ are given by 
\begin{equation}
    l_1=\eta, \quad, l_2=\sqrt{1-\eta^2}-\eta, \quad l_3=2\eta, \quad l_4=l_2, \quad l_5=l_1.
\end{equation}
After one cycle, the parallel coordinate increases by the distance traveled along the closed circuit, which is 
\begin{equation}\label{eq:traveled_distance}
    \mathcal{L}_j = \left\{
    \begin{array}{ll}
      l_1+l_2+l_3k_{3c}^{-3}+l_4+l_5, & \mbox{for} \quad L_1, L_2, L_4, L_5; \\[2pt]
      (l_1+l_2+l_4+l_5)k_{3c}^{3}+l_3,         & \mbox{for} \quad L_3.
    \end{array}\right.
\end{equation}

\section{Numerical solutions for different types of attractors}
\label{app:attractors}
In this section, we consider the numerical solutions obtained for frequencies and aspect ratios
for which the attractor has particular properties. 

In \S~\ref{app:attractor_with_phase_shift}, we consider an attractor 
with a phase shift. 
In \S~\ref{app:attractor_attraction1} and \S~\ref{app:attractor_attraction0}, we consider
an attractor without phase shift but for which the contraction factor is either 1 or 0.  
This type of attractors are obtained on the border of the frequency range of existence 
of a given attractor.  Here we consider the frequencies $\omega_l=0.806225774$ and $\omega_r=0.824949354$ 
which are the limit values for the existence of the equatorial attractor for the aspect ratio $\eta=0.35$  
\citep{rieutordInertialWavesRotating2001}. 

\subsection{Attractor with a phase shift}
\label{app:attractor_with_phase_shift}
\begin{figure}
    \centering
    \begin{subfigure}{0.4\textwidth}
        \centering
        \includegraphics[width=\textwidth]{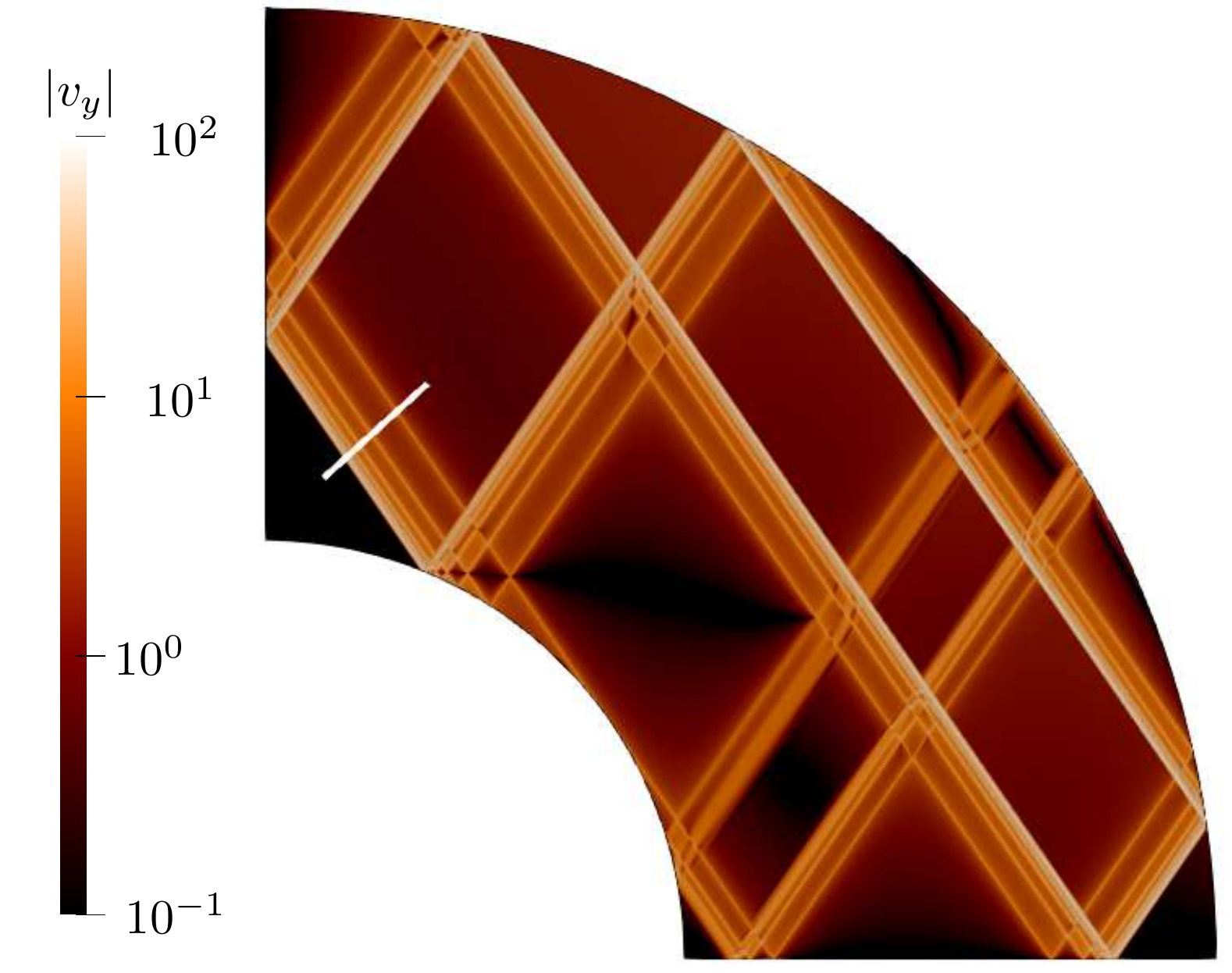}
        \caption{}
    \end{subfigure}%
    \begin{subfigure}{0.6\textwidth}
        \centering
        \includegraphics[width=\textwidth]{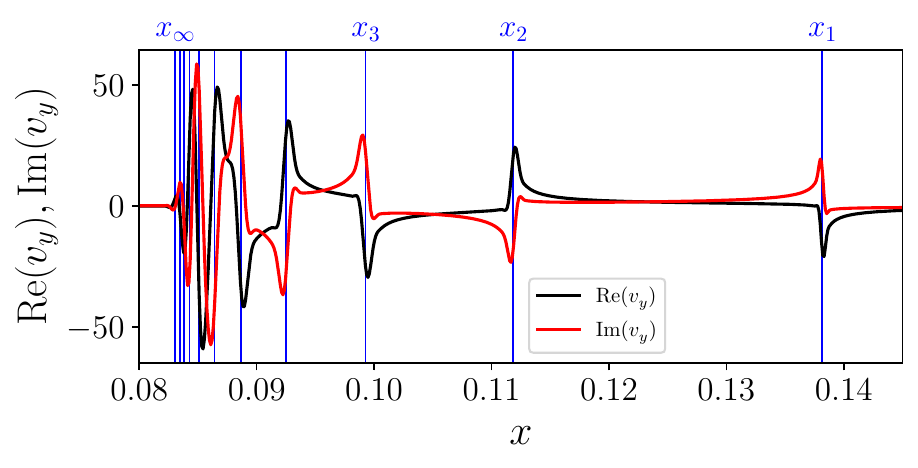}
        \caption{}
    \end{subfigure}
    \vspace{0.5cm}
    \begin{subfigure}{0.45\textwidth}
        \centering
        \includegraphics[width=\textwidth]{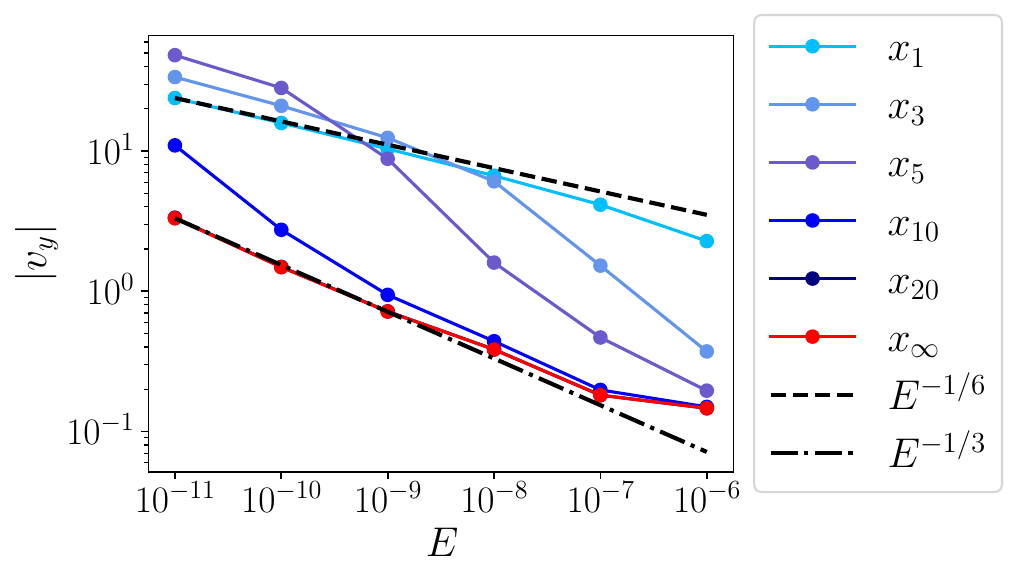}
        \caption{}
    \end{subfigure}
    \caption{Attractors at $\eta=0.44$ and $\omega=1.1329$:
    $(a)$ contour of numerical $|v_y|$ at $E=10^{-11}$;
    $(b)$ profile of $v_y$ at $E=10^{-11}$ on the white cut in $(a)$;
    $(c)$ Ekman number scalings of the velocity amplitude at the critical positions on the white cut in $(a)$.}
    \label{app-fig:attractor_with_phase_shift}
\end{figure}

We can obtain attractors with a phase shift at $\eta=0.44$ and $\omega=1.1329$.
As shown in figure \ref{app-fig:attractor_with_phase_shift}$(a)$, both the polar and equatorial attractors have one touching point on the horizontal axis $Ox$.
Therefore, the phase shift for them is $\pi$.
Figure \ref{app-fig:attractor_with_phase_shift}$(b)$ shows the profile $v_y$ on the white cut in figure \ref{app-fig:attractor_with_phase_shift}$(a)$ at $E=10^{-11}$, which illustrates that the solution at the attractor position is much weaker than those at other critical positions.
Figure \ref{app-fig:attractor_with_phase_shift}$(c)$ demonstrates that the Ekman number scalings of the velocity amplitude are still $E^{-1/6}$ and $E^{-1/3}$ close to the critical point and attractor respectively.

\subsection{Extremely weak attractor without phase shift} 
\label{app:attractor_attraction1}

\begin{figure}
    \centering
    \begin{subfigure}{0.4\textwidth}
        \centering
        \includegraphics[width=\textwidth]{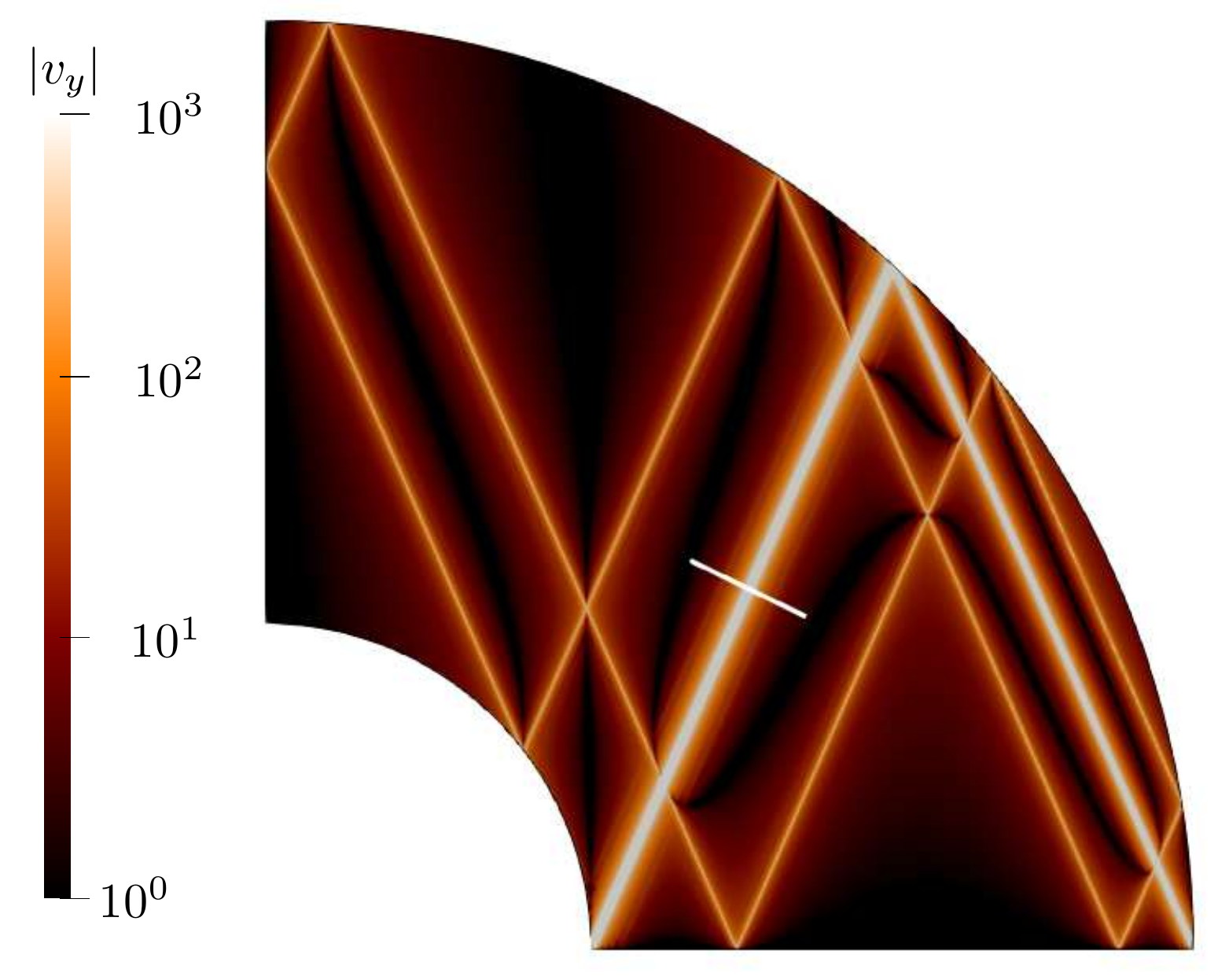}
        \caption{}
    \end{subfigure}%
    \begin{subfigure}{0.6\textwidth}
        \centering
        \includegraphics[width=\textwidth]{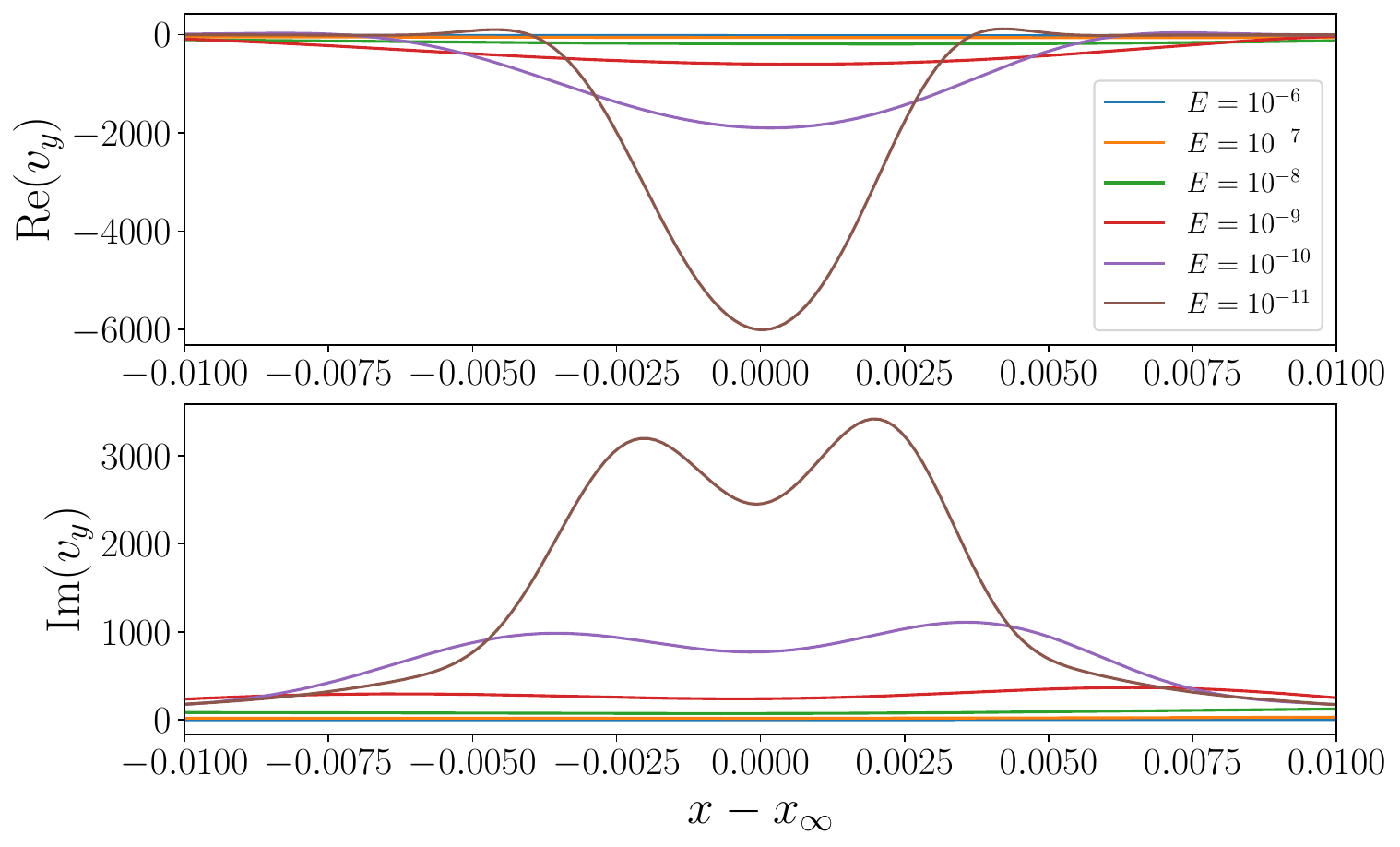}
        \caption{}
    \end{subfigure}
    \vspace{0.5cm}
    \begin{subfigure}{0.35\textwidth}
        \centering
        \includegraphics[width=\textwidth]{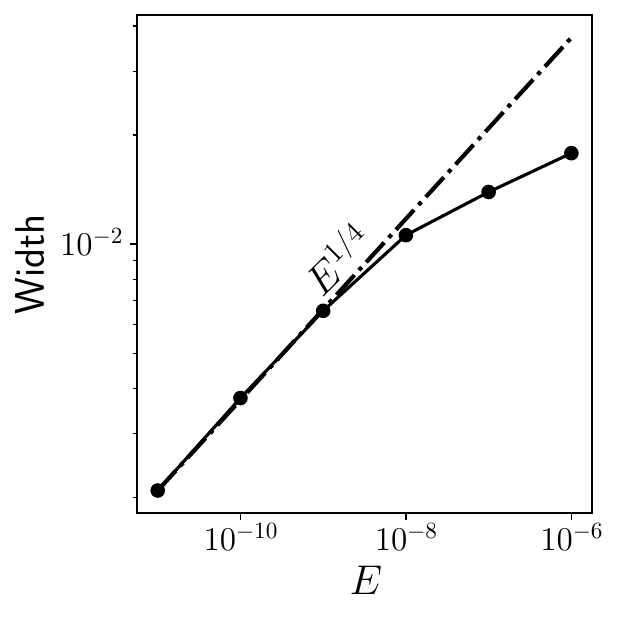}
        \caption{}
    \end{subfigure}%
    \begin{subfigure}{0.35\textwidth}
        \centering
        \includegraphics[width=\textwidth]{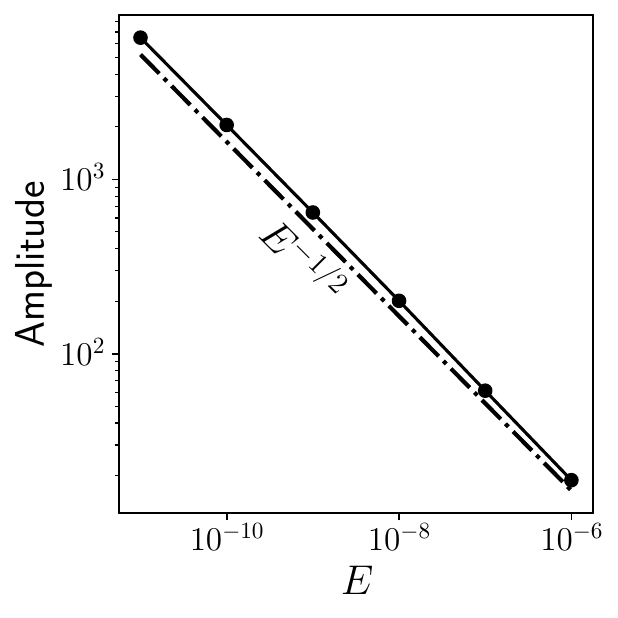}
        \caption{}
    \end{subfigure}
    \caption{Equatorial attractor at $\eta=0.35$ and $\omega=0.806225774$:
    $(a)$ contour of numerical $|v_y|$ at $E=10^{-11}$;
    $(b)$ profiles of $v_y$ on the white cut in $(a)$ at different Ekman numbers ($x_\infty$, attractor position);
    $(c-d)$ Ekman number scalings for the width and velocity amplitude of the shear layer around the equatorial attractor.
    }
    \label{app-fig:attractor_0.8062}
\end{figure}

Figure \ref{app-fig:attractor_0.8062} shows the results for the equatorial attractor at  $\omega=\omega_l$,  for which the contraction factor is close to 1. The equatorial 
attractor is then extremely weak with a Lyapunov number close to $0$. 
For this frequency, the equatorial attractor is  composed of two segments connecting the
 inner core equator with the outer core equator. 

Figure \ref{app-fig:attractor_0.8062}$(a)$ shows that the response around  the equatorial attractor is stronger than around the polar attractor. Its amplitude strongly increases with 
the Ekman number, as observed in figure \ref{app-fig:attractor_0.8062}$(b)$. 
As shown in figure \ref{app-fig:attractor_0.8062}$(c,d)$, the width and the velocity amplitude of the internal shear layer around the equatorial attractor scale as $E^{1/4}$ and $E^{-1/2}$ respectively. This is clearly different from the scalings in $E^{1/3}$ and $E^{-1/3}$ that we have
obtained for an attractor with a contraction factor different from 1. 

Such scalings were already obtained by  \cite{rieutordViscousDissipationTidally2010} 
for attractors with vanishing Lyapunov number forced by tides.
They explained them by a resonance with an attractor eigenmode. The scaling in amplitude 
comes from the frequency of this eigenmode which expands as $\omega \sim \omega_l + E^{1/2} \omega_1$, while the scaling in $E^{1/4}$ of the width is directly related 
to the structure of this eigenmode. 


\subsection{Extremely strong attractor without phase shift}  
\label{app:attractor_attraction0}

\begin{figure}
    \centering
    \begin{subfigure}{0.4\textwidth}
        \centering
        \includegraphics[width=\textwidth]{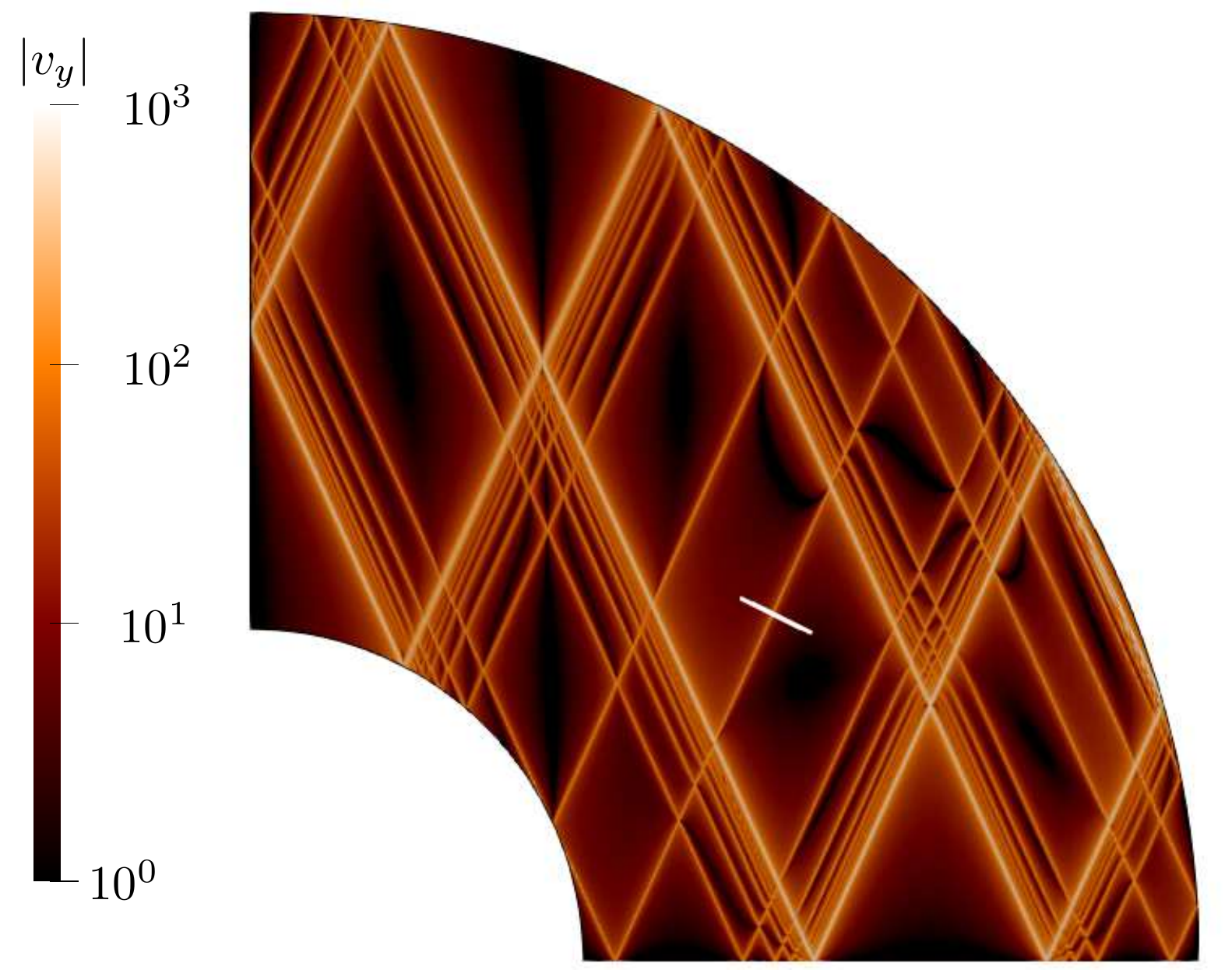}
        \caption{}
    \end{subfigure}%
    \begin{subfigure}{0.6\textwidth}
        \centering
        \includegraphics[width=\textwidth]{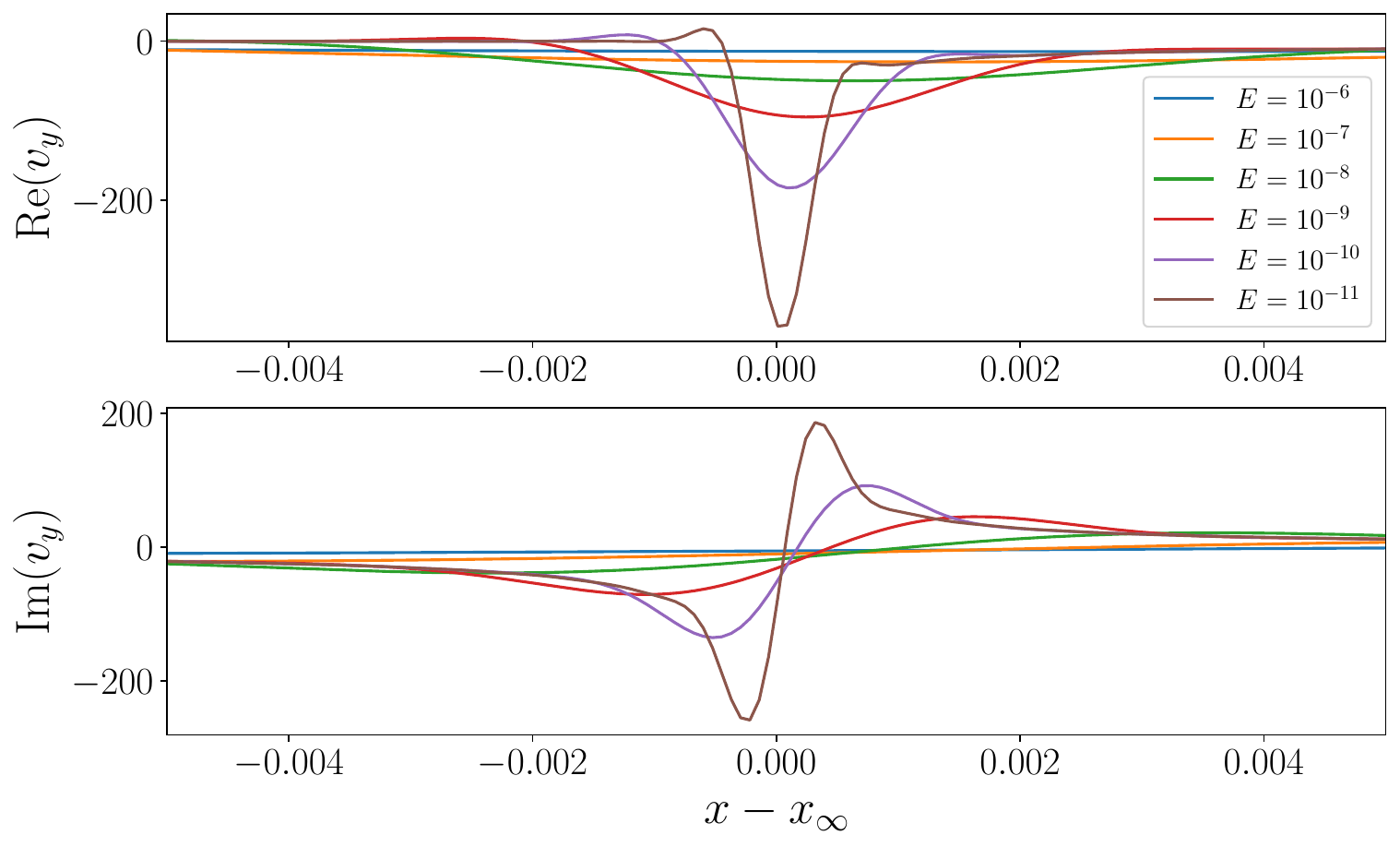}
        \caption{}
    \end{subfigure}
    \vspace{0.5cm}
    \begin{subfigure}{0.35\textwidth}
        \centering
        \includegraphics[width=\textwidth]{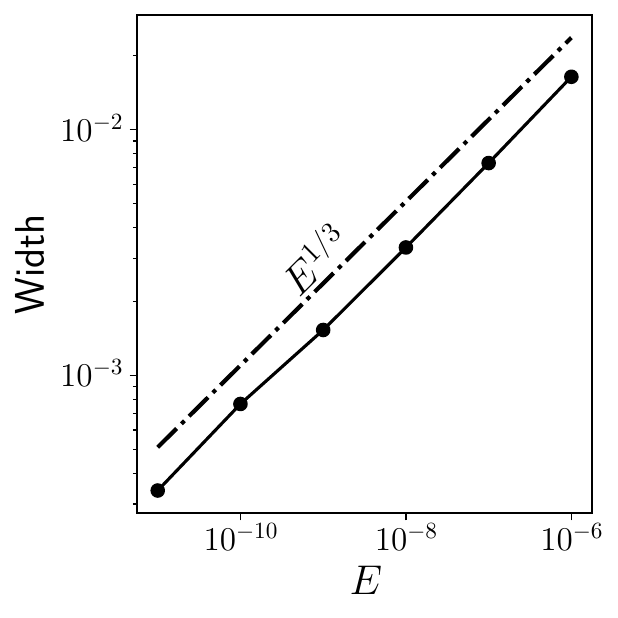}
        \caption{}
    \end{subfigure}%
    \begin{subfigure}{0.35\textwidth}
        \centering
        \includegraphics[width=\textwidth]{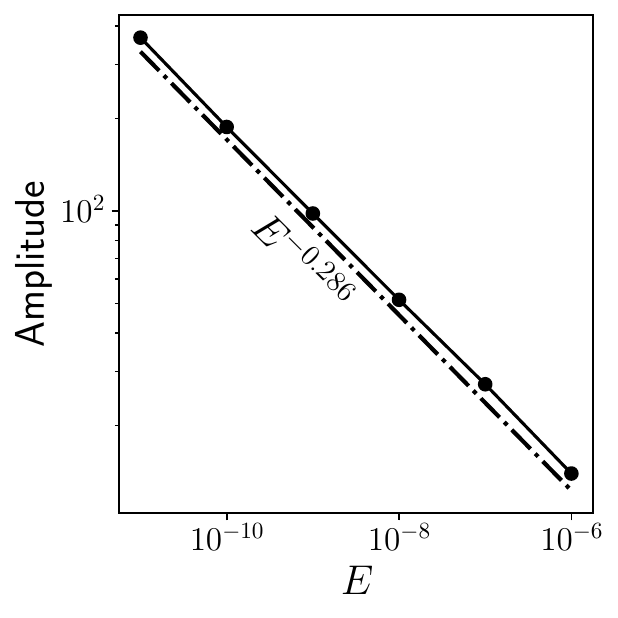}
        \caption{}
    \end{subfigure}
    \caption{Same caption as in figure \ref{app-fig:attractor_0.8062} but for $\eta=0.35$ and $\omega=0.824949354$.}
    \label{app-fig:attractor_0.8249}
\end{figure}

In figure \ref{app-fig:attractor_0.8249},  we show the results for the equatorial attractor at $\omega=\omega_r$ .
For this frequency, the attractor is extremely strong as its Lyapunov number and contraction factor are close to minus infinity and zero respectively.
Figure \ref{app-fig:attractor_0.8249}$(a)$ shows that  the equatorial attractor touches the inner boundary at the critical point. The singularities associated with the critical point and the attractor have then merged in that case.  
The velocity profiles for the internal shear layer around the equatorial attractor are shown in figure \ref{app-fig:attractor_0.8249}$(b)$.
Figure \ref{app-fig:attractor_0.8249}$(c)$ shows that the width of the internal shear layer still scales with $E^{1/3}$.
However, figure \ref{app-fig:attractor_0.8249}$(d)$ shows that the amplitude scaling of the velocity is close to $E^{-0.286}$. This amplitude is smaller than the $E^{-1/3}$ we have found for a regular attractor. 
Surprisingly, the merging of the singularities has therefore not boosted the response.
The amplitude scaling is just in between the 
 $E^{-1/3}$ and $E^{-1/6}$ obtained for the attractor solution and the critical-point solution when
 they are well separated from each other.

\bibliographystyle{jfm}
\bibliography{jfm}

\end{document}